\documentstyle[prl,aps,amsfonts,amssymb,twocolumn,epsfig]{revtex}

\newcommand{\cm}{preprint cond-mat}

\newcommand{\be}{\begin{equation}}
\newcommand{\ee}{\end{equation}}
\newcommand{\bea}{\begin{eqnarray}}
\newcommand{\eea}{\end{eqnarray}}
\newcommand{\ccre}[1]{c^{\dagger}_{{#1}}}

\newcommand{\can}[1]{c^{}_{{#1}}}

\newcommand{\cnum}[2]{c^{\dagger}_{{#1}}c^{}_{{#2}}}

\begin{document}
\title{Low energy fixed points of the $\large\sigma$-$\large\tau$ and
the O(3) symmetric  Anderson models}
\author{R.~Bulla,  A.C.~Hewson and G.-M.\ Zhang   }
\address{
Department of Mathematics, Imperial College,\\
180 Queen's Gate, London SW7 2BZ, United Kingdom\\
}
\draft

\twocolumn[\hsize\textwidth\columnwidth\hsize\csname
@twocolumnfalse\endcsname %

\date{\today} \maketitle
\begin{abstract}
We study the single channel (compactified) models,  the $\sigma$-$\tau$ model 
and the O(3) symmetric  Anderson model, which were introduced by  Coleman {\it et al.}, 
and Coleman and Schofield, as a simplified way to understand the
low energy behaviour of the isotropic and anisotropic two channel Kondo systems.
These models display both Fermi liquid and marginal Fermi liquid behaviour
and an understanding  of the nature of their low energy fixed points may give some general 
insights into the low energy  behaviour of other strongly correlated systems.
We calculate the excitation spectrum at the non-Fermi liquid fixed point of the
$\sigma$-$\tau$ model using conformal field theory, and show that the results are
in agreement with those obtained in
recent numerical renormalization group (NRG) calculations. For the O(3)  Anderson model
we find further logarithmic corrections in the weak coupling perturbation expansion
to those obtained in earlier calculations,
such that the renormalized interaction term now becomes marginally stable rather than
marginally unstable. We derive a Ward identity and a renormalized form of the perturbation
theory that encompasses both the weak and strong coupling regimes and show that the 
 $ \chi/\gamma$ ratio  is 8/3 ($\chi$ is the total susceptibility, spin plus isospin), 
 independent of the interaction $U$ and in agreement with the
NRG calculations.
\end{abstract}
\vskip1.0pc]

\section{ Introduction }
 In developing  a simplified way to understand the
low energy behaviour of the isotropic and anisotropic two channel Kondo systems
   Coleman {\it et al.} \cite{cit}
  and Coleman and Schofield  \cite{cs} introduced two new  single channel 
impurity models.
These  are the $\sigma$-$\tau$ model and the O(3) symmetric  Anderson model.
The $\sigma$-$\tau$ model is similar in form to the 
two channel Kondo model in that a localized spin $s=1/2$ is coupled via an exchange
interaction to two conduction electron terms, but only one of these terms is
the conduction electron spin $\sigma$, the other term is the conduction electron
isospin
$\tau$.
The  Hamiltonian can be expressed in the form,
\begin{eqnarray}
H&=&[J_1\vec{\sigma}(0)+J_2\vec{\tau}(0)]\cdot\vec{s}_d \nonumber \\
& & \hspace{-0.3cm} +
\sum_{n=0,\sigma}^{\infty}t_n(\cnum{n+1,\sigma}{n\sigma}+\cnum{n,\sigma}{n+1,\sigma}),
\label{tsm}
\end{eqnarray}
where $J_1$ and $J_2$ are the two exchange couplings, and the conduction electrons are in
the form of a tight-binding chain with $c^{\dagger}_{n\sigma}$, $c^{}_{n\sigma}$,
the creation and annihilation operators at site $n$, and $t_n$ is the nearest neighbour
hopping matrix element.
The spin operators,
\bea
 \sigma^{+}(n)&=&c^{\dagger}_{n\uparrow}c^{}_{n\downarrow},\quad \sigma^{-}(n)
=c^{\dagger}_{n\downarrow}c^{}_{n\uparrow},\quad 
\nonumber \\ \sigma_{\rm z}(n)&=&{1\over 2}
(c^{\dagger}_{n\uparrow}c^{}_{n\uparrow}-c^{\dagger}_{n\downarrow}c^{}_{n\downarrow}),
\label{spinop}
\eea
 in the basis spanned by the two singly occupied fermion states, 
 $|0(\uparrow), 1(\downarrow)\rangle$ and
$|1(\uparrow), 0(\downarrow)\rangle$, give a representation for the SU(2)
 algebra for spin $\sigma={1\over 2}$. The
isospin operators,
$$\tau^{+}(n)=(-1)^nc^{\dagger}_{n\uparrow}c^{\dagger}_{n\downarrow},
\quad \tau^{-}(n)=(-1)^nc^{}_{n\downarrow}c^{}_{n\uparrow},$$
\be \tau_{\rm z}(n)={1\over 2}
(c^{\dagger}_{n\uparrow}c^{}_{n\uparrow}+c^{\dagger}_{n\downarrow}
c^{}_{n\downarrow}-1),\label{isospinop}\ee
 give a  representation of the same algebra in the space spanned
 by the zero and double occupation states, $| 0,0\rangle$ and $| 1,1\rangle$.
The operators for the two conduction electron channels of the two channel Kondo model 
have been 
replaced by the operators of  a {\em single} conduction electron channel, so the term 
`compactified'
has also been used to describe the model.  The model as introduced by Coleman {\it et al.} was
for the isotropic case $J_1=J_2$.\par
The O(3) symmetric Anderson model is a modified form of the symmetric 
 Anderson model (SAM).
The symmetric  Anderson model can be written in the form,
\bea H&=&\sum_{n=0,\sigma}^{\infty}
t_n(\cnum{n+1,\sigma}{n\sigma}+\cnum{n\sigma}{n+1,\sigma}) \nonumber \\
 &&\hspace{-0.5cm}+ \sum_{\sigma}V(
\cnum{0\sigma}{d\sigma}+\cnum{d\sigma}{0\sigma})+U(n_{d\uparrow}-{1\over 2})
(n_{d\downarrow}-{1\over 2}),\label{sam}\eea
where $\ccre{d\sigma}$ and $\can{d\sigma}$ are creation and annihilation operators 
for the localized impurity  d state with spin $\sigma$. 
 The electrons in the impurity state interact
via a Coulomb matrix element $U$, and $V$ is the hybridization matrix element
of the impurity state with a conduction electron chain which is of the same form
as that given in ({\ref{tsm}).
 This model was modified by Coleman and Schofield \cite{cs}
by the addition of an anomalous hybridization term,
\be -V_{\rm a}\{\ccre{d\downarrow}\ccre{0\downarrow}+\can{0\downarrow}\can{d\downarrow}+
\ccre{d\downarrow}\can{0\downarrow}+\ccre{0\downarrow}\can{d\downarrow}\}
\label{ah}.\ee
The motivation for the addition of such a term is that if one applies a Schrieffer-Wolff
transformation to the model in the local moment regime (large $U$) then 
it maps into the $\sigma$-$\tau$ model with $J_1$ and $J_2$ given by
\be J_1={{4V(V-V_{\rm a})}\over{U}},\quad  J_2={{4VV_{\rm a}}
\over{U}}.\ee
The symmetry of the model can be made explicit by the introduction of the Majorana
fermion representation.
 The Majorana
fermion operators for the d electrons are defined by
\begin{eqnarray}
d_1={1\over {\sqrt 2}}(c^{}_{d\uparrow}
+c^{\dagger}_{d\uparrow})&\quad& d_2={i\over {\sqrt 2}}(c^{}_{d\uparrow}
-c^{\dagger}_{d\uparrow}),
\nonumber \\
d_3={-1\over {\sqrt 2}}(c^{}_{d\downarrow}
+c^{\dagger}_{d\downarrow})&\quad&  d_0={i\over {\sqrt 2}}
(c^{}_{d\downarrow}-c^{\dagger}_{d\downarrow}),
\end{eqnarray}
which satisfy the commutation relations,
\begin{equation}\{d_{\alpha}, d_{\beta}\}=
\delta_{\alpha,\beta},\label{cr}\end{equation}
where $\{\}$ indicates an anticommutator.\par
The Majorana operators for the conduction electrons are similarly defined by
\begin{eqnarray}
\psi_1(n)&=&{e^{in\pi/2}\over {\sqrt 2}}((-1)^nc^{}_{n\uparrow}
+c^{\dagger}_{n\uparrow}), \nonumber \\
\psi_2(n)&=&{ie^{in\pi/2}\over {\sqrt 2}}((-1)^nc^{}_{n\uparrow}
-c^{\dagger}_{n\uparrow}), \nonumber \\
\psi_3(n)&=&{-e^{in\pi/2}\over {\sqrt 2}}((-1)^nc^{}_{n\downarrow}
+c^{\dagger}_{n\downarrow}) \nonumber \\ 
\psi_0(n)&=&{ie^{in\pi/2}\over {\sqrt 2}}
((-1)^nc^{}_{n\downarrow}-c^{\dagger}_{n\downarrow}),
\end{eqnarray}
with commutation relations as in (\ref{cr}).
The modified Anderson model in
this  representation  takes the form      
\bea
H&=&i\sum_{\alpha=1}^3 V\psi_{\alpha}(0)d_{\alpha}
+iV_0\psi_{0}(0)d_{0} \nonumber \\
&+& i\sum_{\alpha=0}^3 \sum_{n=0}^{\infty}t_n\psi_{\alpha}(n+1)
\psi_{\alpha}(n)+Ud_{1}d_{2}d_{3}d_{0},\label{hamm}
\eea
where $V_0=V-2V_{\rm a}$.  
The Hamiltonian is invariant under the orthogonal
transformation,
\be d_{\alpha}=\sum_{\beta=1}^3 a_{\alpha, \beta} d_{\beta}, \quad \psi_{\alpha}(n)=
\sum_{\beta=1}^3 a_{\alpha, \beta} \psi_{\beta}(n), \ee
where $\sum_{\beta} a_{\alpha, \beta}a_{\alpha^{\prime}, \beta}=\delta_{\alpha,\alpha^{\prime}}$
and $\alpha, \alpha^{\prime},\beta=1,2,3$,
which corresponds to O(3)  symmetry as the 0 Majorana states are not included. For
$V_0=V$  we recover the standard symmetric Anderson model ($V_{\rm a}=0$)
and the model is then symmetric under the corresponding O(4) transformation
with the 0 Majorana states included.
\par
 These two models do not conserve either the number of particles or the spin. However,
there is a combination of spin plus isospin, which we denote by $\vec{j}_{\rm tot}$, which is conserved. 
For the Anderson model this corresponds
to 
\be \vec{j}_{\rm tot}= \vec{s}_d + \vec{\tau}_d +\sum_{n=0}^{\infty}(\vec{\sigma}_n +\vec{\tau}_n),\ee
where $\vec{\tau}_d$ is the isopin operator for the impurity site. As the $\sigma$-$\tau$ model
corresponds to the localized limit in which the impurity charge fluctuations are suppressed
the $\vec{\tau}_d$ is omitted for this model. Clearly this combination satisfies the usual
SU(2) angular momentum commutation relations, so the many-body states of these models can be
classified by the quantum numbers $j_{\rm tot}$ and $j_{{\rm tot},z}$. The combination of spin and isospin for a
particular site $\vec{j}_n$ when expressed in terms of the Majorana fermions takes the form,
\be \vec {j}_{n}=\vec{\sigma}_n+\vec{\tau}_n={{-i}\over{2}}\vec{\psi}(n) \times \vec{\psi}(n),
\label{deft}\ee
where $\vec{\psi}(n)=(\psi_1(n),\psi_2(n),\psi_3(n))$. Hence we will refer to the $\alpha=1,2,3$
Majorana components as the components of a vector field, and the $\alpha=0$ as the scalar 
component.
\par
The $\sigma$-$\tau$ model differs from the two channel Kondo model in that there is no
 possibility of
overscreening the impurity spin. Overscreening occurs in the two channel case in the strong 
coupling limit 
 $J_1\to\infty$, $J_2\to\infty$, when a conduction electron in each of the two channels binds to 
the impurity spin resulting in a localized state at the impurity which again has spin $1/2$.
In the $J_1\to\infty$, $J_2\to\infty$ limit of the $\sigma$-$\tau$ model, however, because the spin
and isospin configurations are mutually exclusive states of the same channel the impurity cannot
bind to both simultaneously and overcompensation cannot occur. 
\par
Coleman {\it et al.} established a
correspondence between the two models in perturbation theory (after bosonization) and they 
conjecture that the
low energy physics of the two models is essentially the same (see also \cite{sch}),  with the extra states of the two
channel model playing no significant role. For the isotropic case, $J_1=J_2=J$ they obtained
an effective Hamiltonian for the $\sigma$-$\tau$ model by assuming that the low energy fixed
point corresponds to $J=\infty$, and then calculating the leading order terms in an expansion
in $D/J$ where $D$ is the band width of the conduction electrons. The effective Hamiltonian to
leading order in $D/J$ has an interaction term which corresponds to that in the O(3) Anderson
 model
with $U\sim D^3/J^2$ and $V_0=0$,
so that corrections to the fixed point correspond to a low order
 perturbation expansion in powers of $U$ for the O(3) Anderson model. The second order terms
generate  ${\rm ln}T$ terms in the susceptibility and specific heat, and
 a Wilson ratio for the logarithmic terms of $8/3$,
 which is the low energy behaviour of the two channel Kondo model.\par
There are several reasons why we think these models are worthy of further study. The low
order perturbation theory in $U$ (for $V_0=0$) gives a self-energy corresponding to the
marginal Fermi liquid form whose imaginary part is given by
\be {\rm Im}\Sigma^{(2)}(\omega,T)=-{\pi\over 2}\left({U\over{\pi\Delta}}\right)^2
|\omega|{\rm coth}\left({|\omega|\over{2T}}\right),\ee
and real part,
\be {\rm Re}\Sigma^{(2)}(\omega,T)\approx\left({U\over{\pi\Delta}}
\right)^2\omega{\rm ln}\left({x\over\Delta}\right),\ee
where  $x={\rm max}(|\omega|,T)$, $\Delta=\pi V^2 \rho_c(0)$ and $\rho_c(0)$ is the 
density of conduction electron states at the Fermi level \cite{zh,cit}.\par
Marginal Fermi liquid theory was put forward in a phenomenological interpretation of the 
behaviour of the normal state of the high $T_{c}$ superconductors so it is of some interest to
understand fully a microscopic model which displays this behaviour \cite{vlsa}. 
In higher order perturbation theory diagrams were found that give logarithmic contributions 
 to the irreducible four vertex (at zero frequency)
with a sign indicating that the weak coupling fixed point 
\cite{zh} might be unstable.
 It is important to understand the role of these terms in
the low energy behaviour of the model, and to determine whether or not the marginal Fermi liquid
behaviour corresponds to a stable fixed point. Apart from the non-Fermi liquid behaviour of
the isotropic model ($J_1=J_2$) it has been also conjectured, based on Bethe ansatz 
calculations, that the behaviour of the anisotropic two channel model also differs from that 
for a Fermi liquid fixed point \cite{aj}. If the $\sigma$-$\tau$ model and the two channel Kondo model
 have the
same low energy fixed point then  the anisotropic $\sigma$-$\tau$
should behave in a similar way. These are points that need further clarification.\par

In the derivation of the low energy behaviour of the $\sigma$-$\tau$ model
 based on the $1/J$
expansion there are points that are not clear. For the single channel Kondo model it has been
 shown by Nozi\'eres \cite{noz} that, 
 though the low energy  fixed point corresponds to $J=\infty$,
the low energy physics cannot be calculated from a $1/J$ expansion (which gives a non universal
Wilson ratio).  We know that for this model there is only one energy
scale which is the Kondo temperature
$T_{\rm K}$, and so the Wilson ratio ($R$) has a universal value ($R=2$).
The effective Hamiltonian that determines the low energy impurity behaviour depends only 
on $T_{\rm K}$, 
and asymptotic expansions about the fixed point are in powers of 
 $T/T_{\rm K}$ (or frequency or magnetic field compared to the Kondo temperature).
The leading interaction terms in the effective Hamiltonian generated by a $1/J$ expansion
  do satisfy the Wilson criteria 
 for the
leading corrections to the fixed point because (a) they involve operators local to the 
impurity, and
(b) they are consistent with the symmetries of the model. However the  combination of these
terms generated  in this effective Hamiltonian ($D/J\ll 1$), does not correspond to the 
correct low energy effective 
Hamiltonian for the Kondo model. The $1/J$ effective Hamiltonian, as to be expected, reproduces 
the $D/J\ll 1$ physics of the model with a non-universal Wilson ratio dependent upon $D/J$,
and so does not give the correct low energy behaviour about the fixed point of the weak 
coupling model.
The applicability  of the $1/J$ expansion to derive the excitations about the fixed point of
the weak coupling $\sigma$-$\tau$ model needs to be clarified.\par

In this paper we bring together information on the low energy behaviour of these models, obtained
from  different many-body techniques, conformal field theory, numerical renormalization group 
and perturbation theory in order to clarify and resolve some of these outstanding 
issues. We first of all consider the nature of the excitations in a free single component 
Majorana chain with various
types of boundary conditions, and then calculate possible forms of the  many-body excitation 
spectra
when the component systems are put together. We relate these results to those obtained
using  conformal field theory. The conformal field approach can also be used
to calculate the many-body excitations of the isotropic $\sigma$-$\tau$ model for a particular 
value of the exchange interaction $J=J^*$, corresponding to strong coupling. We  show that the
excitations of the model at this value of $J$ correspond to those found at the fixed point in
 recent
numerical renormalization group calculations \cite{bh}. From the conformal field theory we also obtain the
form of the operators which give the leading order corrections to this fixed point.\par
For the O(3) Anderson model we find additional logarithmic terms 
to those found in earlier work \cite{zh}, and when these are included the 
renormalized interaction
is  marginally irrelevant rather than marginally relevant so the marginal fixed
point is stable. We also derive a Ward
identity from which we can deduce the Wilson ratio for the total susceptibility (spin plus isospin)
of the impurity to the specific heat coefficient. 
We show that a  renormalized perturbation expansion for the Fermi liquid fixed point ($V_0\ne 0$)
can be generalized for the marginal Fermi liquid fixed point ($V_0=0$). 
The renormalized perturbation theory encompasses both the strong and weak coupling limits.
Finally we relate our results to those obtained by other methods, and to the results for
the two channel Kondo model.\par

\par
\section{Free Majorana Fermion Chain}\par
\bigskip
 
A Fermi liquid fixed 
point of an interacting system is characterized by quasiparticle excitations  which are in
 one-to-one correspondence with the single particle excitations of the non-interacting system.
In the models we are considering here the single particle excitations of 
the non-interacting system are independent Majorana fermion excitations. As a preliminary
to investigating the fixed point of the interacting systems we look first of all
 at the possible excitations of a free Majorana chain.  We consider just a one component 
($\alpha=0$)
 chain with $N$ sites described by the Hamiltonian,

\begin{equation}
  H = it \sum_{n=0}^{N-1} {\psi}_{0}(n) {\psi}_{0}(n+1), \label{eq:ham}
\end{equation}
where we have taken the  hopping matrix-element $t$ (real) to be the same
 between all neighbouring sites.
The number
of sites is assumed to be even and we use periodic or antiperiodic boundary
conditions ${\psi}_0(N)\equiv \pm {\psi}_0(0)$.

The Fourier-transformed operators are defined as
\begin{equation}
 \bar  \psi_{0}(k) = \frac{1}{\sqrt{N}} \sum_{n=0}^{N-1} e^{i\frac{2\pi}{N}kn} 
{\psi}_{0}(n),
\label{eq:psi_k}
\end{equation}

\par

We consider first the case of periodic boundary conditions.
In this case  $k$  takes only the integer
values, 
\begin{equation}
   k = -\frac{N}{2}+1, \ldots , -1,0,1, \ldots, \frac{N}{2}.
\end{equation}
The Hamiltonian (\ref{eq:ham}) can be diagonalized by using Eq.\ 
(\ref{eq:psi_k}).
\begin{eqnarray}
  H &=& \sum_{k<0} it e^{i\frac{2\pi}{N}k} \bar \psi_0(k) \bar\psi_0(-k) 
    + \sum_{k>0} it e^{i\frac{2\pi}{N}k} \bar \psi_0(k)\bar \psi_0(-k) \nonumber \\
  &+& it\bar \psi_0(0)\bar \psi_0(0) - it\bar \psi_0(\frac{N}{2})\bar \psi_0(-\frac{N}{2}). 
 \label{eq:HII}
\end{eqnarray}
Due to $\bar\psi_0(-\frac{N}{2}) =\bar \psi_0(\frac{N}{2})$, the last two
terms cancel. There is a Majorana fermion state
at the Fermi level but it does not appear explicitly in the Hamiltonian as we take $\mu=0$.
For $k>0$ we define the operators
\begin{equation}
 \bar \psi^\dagger_0(k) \equiv \bar\psi_0(-k).
\end{equation}
Using Eq.\ (\ref{eq:psi_k}) and the commutation relation for the
Majorana fermions $\psi_0(n)$, one obtains
\begin{equation}
   \left\{\bar \psi_0(k) ,\bar\psi^\dagger_0(k^\prime) \right\}_+ = 
   \delta_{k,k^\prime}
\end{equation}
which is just the anticommutation relation for ordinary fermion operators.
That means that we can treat the operators $\bar\psi^\dagger_0(k)$
and $\bar\psi_0(k)$ for $k>0$ as fermion operators in the following discussion.
The Hamiltonian now takes the simple form
\begin{equation}
  H=\sum_{k>0} \epsilon_{k}\bar \psi_0^\dagger(k)\bar \psi_0(k)
   + E_{g,{\rm p}} ,
\end{equation}
with the dispersion 
\begin{equation}
  \epsilon_{k} = 2t \sin \frac{2\pi}{N}k  \label{eq:disp}
\end{equation}
and the ground state energy 
\begin{equation}
E_{g,{\rm p}} = it \sum_{k>0} e^{i\frac{2\pi}{N}k}.
\end{equation}
In the limit of large $N$, the single particle excitation energies
$\epsilon_{k}$ are given by
\begin{equation}
  \epsilon_{k} = \frac{4t \pi}{N}k = \frac{\pi v_{\rm F}}{l} k ,
  \ \ k=1,2,3,\ldots .
\end{equation}
for small $k/N$. There is a similar linear branch for $k/N\sim {1\over 2}$.
The former we will refer to as the left branch and the latter the right branch.
Here we have expressed $t$ in terms of the Fermi velocity $v_{\rm F}=2t$
and replaced the number of sites $N$ by the length of the chain
$L=N=2l$ (we introduce $l$ to be consistent with the notation used in Sec.\ III).

The ground state energy is given by
\begin{equation}
E_{g,{\rm p}} = -t \frac{\cos(\frac{\pi}{N})}{\sin(\frac{\pi}{N})}
   \approx
   -\frac{v_{\rm F}l}{\pi} \left(1-\frac{1}{12} \frac{\pi^2}{l^2}\right).
\end{equation}
In the case of antiperiodic boundary conditions, $k$ can only take the 
half-integer values
\begin{equation}
   k = -\frac{N}{2} +\frac{1}{2},\ldots,-\frac{1}{2},
   \frac{1}{2}, \ldots, \frac{N}{2} -\frac{1}{2}.
\end{equation}
In contrast to Eq.\ (\ref{eq:HII}) there is no degree of freedom
at the Fermi level, so the Hamiltonian reads
\begin{equation}
  H = \sum_{k=1/2}^{(N-1)/2} \epsilon_{k}\bar \psi_0^\dagger(k)\bar \psi_0(k)
   + E_{g,{\rm a}}\label{mc}
\end{equation}
with the dispersion Eq.\ (\ref{eq:disp}) and the ground state energy
\begin{equation}
  {E}_{g,a} = it \sum_{k>0} e^{i\frac{2\pi}{N}k}.
\end{equation}
In the limit of large $N$, the single particle excitation energies
 are given by
\begin{equation}
  \epsilon_{k} =  \frac{\pi v_{\rm F}}{l} k ,
  \ \ k=\frac{1}{2},\frac{3}{2},\frac{5}{2},\ldots .
\end{equation}
For the ground state energy we find
\begin{equation}
  E_{g,a} = -t \frac{1}{\sin(\frac{\pi}{N})}
   \approx -\frac{v_{\rm F}l}{\pi} 
    \left(1+\frac{1}{24} \frac{\pi^2}{l^2}\right).
\end{equation}
The difference between the ground state energies for
 periodic and antiperiodic boundary conditions is
\begin{equation}
   E_{g,p}-E_{g,a} =  -\frac{v_{\rm F}l}{\pi}\frac{\pi^2}{l^2}
     \left(-\frac{1}{12} -\frac{1}{24}\right) = \frac{1}{8} 
   \frac{\pi v_{\rm F}}{l} ,
\end{equation}
where both linearized branches of (\ref{mc}) have been taken into account.
With the restriction to only one branch of Majorana fermions,
the single particle spectrum is given by Fig.\ 1(a) and 1(b) for
periodic and antiperiodic boundary conditions. \par

From the single particle spectrum we can also build up the spectrum for
the many-body excitations  (relative to the ground state) of 
the scalar Majorana fermion chain, which corresponds to
progressively creating 'particles' in the states with  $k\ge 0$. The ground
state is a vacuum state with no particles. The 'particles' which are created
are combinations of the usual particle and hole excitations  of a free electron
system, which are degenerate as a result of the particle-hole symmetry.
 The resulting many-body energies $E_{\rm ex}$
 and their degeneracies $dg$ are given in
Tables I and II, where $n_k$ denotes the occupation of the single particle state of
energy $\epsilon_k$.

We can build up the many-body excitations of the vector Majorana fermion chain
corresponding to the Hamiltonian,
\begin{equation}
  H = it \sum_{n=0}^{N-1}\sum_{\alpha =1}^3 {\psi}_\alpha(n) 
      {\psi}_\alpha(n+1), \label{eq:ham_vec}
\end{equation}
in a similar way.
The single particle spectra for this model for periodic and antiperiodic
boundary conditions are given in Fig.\ 2(a) and 2(b), respectively.
The number of excited states per energy level is simply multiplied by
three as compared to Fig.\ 1.
The resulting many-body states and their degeneracies are given in Table III
and IV.

\section{Conformal Field Theory (CFT)}\par
In this section, we apply a current algebra approach to the  channel isotropic $\sigma$-$\tau$
model along the lines developed by Affleck and Ludwig \cite{al} 
for the two channel Kondo model. In our case, however, a new separation scheme is used 
in which  the 
conduction electrons, expressed as Majorana fermions, are separated into a 
 density corresponding to total spin (spin plus isospin, the vector channel) and a 
single (scalar)
 component Majorana fermion.
 As in the other examples where this technique is used the model will be
solved at a particular  finite but large value of the coupling $J^*$. At this value of
the coupling  the impurity spin
can be  absorbed into the conduction electrons, and the algebra which
determines the excitation spectrum is then of the same form as that for
the non-interacting  system. \par
 The $\sigma$-$\tau$ Hamiltonian
can be written as
$$ H=it\sum_{n=0}^{\infty}\sum_{\alpha=0}^3
    \psi_{\alpha}(n+1)\psi_{\alpha}(n)
  +J \vec{j}_0\cdot\vec{s}_d.$$
where $\vec{j}_n$ is the combination of spin and isospin as defined earlier
 in Eq.\ (\ref{deft}).  In this form the vector and scalar Majorana fermion
terms are decoupled so that the model can be
divided into two parts (see Fig.\ 3)  
\begin{eqnarray}
&& H_{\rm sc}=it\sum_{n=0}^{\infty}\psi_0(n+1)\psi_0(n), \nonumber \\
&& H_{\rm vec}=it\sum_{n=0}^{\infty}\sum_{\alpha=1}^3
               \psi_{\alpha}(n+1)\psi_{\alpha}(n)+J\vec{j}_0\cdot\vec{s}_d.
\end{eqnarray}
If the lattice spacing is $a$ we can define a spatial coordinate $x=na$
for each spin and in the limit as $a\to 0$ we use Eqs.\ (\ref{spinop}) and (\ref{isospinop}) 
to define
both continuum spin and isospin densities with components, $\sigma^{\gamma}(x)$
and $\tau^{\gamma}(x)$, where $\gamma=x,y,z$.
These densities satisfy a level $k=1$ $SU(2)$ Kac-Moody algebra:
\begin{eqnarray}
&& [s^{\alpha}(x),s^{\beta}(x')] \nonumber \\ &&
  =i\epsilon^{\alpha\beta\gamma}s^{\gamma}(x)\delta(x-x')
  + \frac{ik}{2\pi}\delta_{\alpha\beta}\delta^{\prime}(x-x').
\end{eqnarray}
where $s^{\gamma}(x)=\sigma^{\gamma}(x),\tau^{\gamma}(x)$, and
$\delta^{\prime}$ is  the  derivative of the delta function.
If we  define similarly a density $\vec{j}(x)$ for the combined spin 
and isospin then the three components of this density satisfy
 an $SU(2)$ level $k=2$ Kac-Moody algebra.
\par

In the continuum limit, the linearization of the dispersion relation can be 
made, and the conduction electrons can be expressed in terms of left- and 
right-moving Majorana fermions satisfying the  conditions 
$ \psi_{\alpha,L}(0)=\psi_{\alpha,R}(0) $ with $\alpha=0,1,2,3$. In fact, 
one can continue the Majorana fermions to the negative $x$-direction by defining
$ \psi_{\alpha,L}(x)=\psi_{\alpha,R}(-x) $, so the Hamiltonians can now be 
written in terms of left-moving fermions only on the full $x$-axis. The Hamiltonian
that describes the excitations for the scalar 
part  becomes 
\begin{equation}
 H_{\rm sc}=\frac{v_{\rm F}}{2\pi}\int_{-\infty}^{\infty}dx
:\psi_0(x)i\partial_x\psi_0(x):,
\end{equation}
where the subscripts "L" have been omitted, the Fermi velocity is $v_{\rm F}=2ta$, and
$ :\hspace{.3cm}: $ indicates normal ordering. This Hamiltonian has the
same form as  the one which describes the two dimensional Ising model at the
critical point
\cite{car}.\par
 For the vector part it is possible to apply the standard technique of conformal 
field theory and write the non-interacting part in the Sugawara form \cite{kz}, which is
quadratic
in the densities rather than the fields,  so that $H_{\rm vec}$ takes the form,  
\begin{equation}
 H_{\rm vec}= \frac{v_{\rm F}}{2\pi}\int_{-\infty}^{\infty}dx
             \left[ \frac{1}{4} :\vec{j}(x)\cdot\vec{j}(x): 
               +\frac{2J\pi}{v_{\rm F}}\delta(x)\vec{j}(x)\cdot\vec{s}_d \right],
\end{equation}
which is normal ordered with respect to the non-inter\-acting system. This
 Hamiltonian is expressed entirely in terms
of the combined spin density $\vec{j}(x)$ and the impurity spin, and is the only part
which contains the interaction term.
 
We choose boundary conditions $\psi_{\alpha}(l)=e^{2i\delta}\psi_{\alpha}(-l)$, on
a large circle $ -l\leq x\leq l $ for
$ \alpha=0,1,2,3 $. The  momenta are then given by
$k_m=\frac{\pi}{l}\left(m+\frac{\delta}{\pi}\right)$ and the 
energy levels with respect to the chemical potential are expressed by 
$E_m=v_{\rm F}k_m$ with $ m=0,\pm 1,\pm 2,\cdot\cdot\cdot$. 
In the case of particle-hole symmetry, $\delta=0$ corresponds to the case of 
odd number of sites ($ m\in \bf{Z}$, integers) in the lattice version, while 
$\delta=\pi/2$ to the case of even number sites ($ m\in\bf{Z}+\frac{1}{2}$, 
half-odd integers), these correspond to 
periodic boundary condition (PBC) and anti-periodic boundary condition (APBC),
respectively.\par

We now look at the $J=0$ case with APBC and PBC.
After taking a Fourier transform, the scalar part becomes 
\begin{eqnarray}
 H_{\rm sc}^{(0)} = && \frac{\pi v_{\rm F}}{l}
  \sum_{m>0}(m+\frac{1}{2})\psi_0(-m)\psi_0(m)
\nonumber \\ &&
 +\left \{
 \begin{array}{ll}
  -\frac{\pi v_{\rm F}}{l}\frac{c_{\rm sc}}{24}, & m\in \bf{Z}+\frac{1}{2} \\
  \frac{\pi v_{\rm F}}{l}\frac{c_{\rm sc}}{12}, & m\in \bf{Z}
 \end{array} \right.\end{eqnarray}
where $c_{\rm sc}={1\over 2}$ is the central charge of the Virasoro algebra,
\be  [L_n,L_m]=(n-m)L_{n+m}+{c\over 12}(n^3-n)\delta_{n+m,0}      , \ee
where $L_n$ is the nth component of the mode expansion of the stress-energy
tensor,
which can be used to generate the excitations of the system (for details see \cite{gin}).
The eigenvalues of this Hamiltonian can be expressed in the form 
$$ E_{\rm sc}=\frac{\pi v_{\rm F}}{l}\left[-\frac{1}{48}+\Delta_{\rm Ising}
           +n_{\rm sc}\right] ,$$
where $\Delta_{\rm Ising}=0$, $\frac{1}{2}$, $\frac{1}{16}$. This notation 
is that used in the context of the two dimensional Ising model where the values
of $\Delta_{\rm Ising}$ correspond to the  scaling 
dimensions of the primary fields: identity operator $I$, energy density 
operator $\epsilon$, and Ising order parameter $\sigma$, respectively. For
APBC, the first two primary fields appear and each with degeneracy 
$dg_{\rm sc}=1$. In the PBC case, only $\sigma$ appears with a degeneracy 
$dg_{\rm sc}=2$.  These results agree with those obtained in Tables I and II
in the previous section. 
Those in Table I correspond to $\Delta_{\rm Ising}={1\over 16}$,
 and in Table II correspond to $\Delta_{\rm Ising}=0, {1\over 2}$.
\par
In terms of the Fourier transform of the spin plus isospin density,  
$$\vec{j}(m)=\frac{1}{2\pi}\int_{-l}^{l}dx e^{im\pi x/l}\vec{j}(x),$$
 the free vector part may be cast 
into the form,
\begin{eqnarray}
 H_{\rm vec}^{(0)} 
 = && \frac{\pi v_{\rm F}}{2l}\sum_{m=1}^{\infty}\vec{j}(-m)\cdot\vec{j}(m) 
 \nonumber \\ && 
  \hspace{-0.9cm}
  +\frac{\pi v_{\rm F}}{l}\frac{1}{4}\vec{j}(0)\cdot\vec{j}(0)
  +\left \{
  \begin{array}{ll}
   -\frac{\pi v_{\rm F}}{l}\frac{c_{\rm vec}}{24}, & m\in \bf{Z}+\frac{1}{2} \\
   \frac{\pi v_{\rm F}}{l}\frac{c_{\rm vec}}{12}, & m\in \bf{Z}
  \end{array} \right.
\end{eqnarray}
where $c_{\rm vec}={3\over 2}$ is the central charge for the vector field.
The eigenvalues for this part of the Hamiltonian which describes
 the vector Majorana fermions can be written in the form,
$$ E_{\rm vec}=\frac{\pi v_{\rm F}}{l}\left[-\frac{1}{16}+
               \frac{j(j+1)}{4}+n_{\rm vec}\right] ,$$
where $j$ is a good quantum number and $n_{\rm vec}$ is a non-negative integer.
The primary fields of the conformal field theory corresponding
to this  model are $j=0$ (singlet), $j=\frac{1}{2}$ (doublet), and $j=1$ 
(triplet), with scaling dimension $j(j+1)/4$ and degeneracy 
$dg_{\rm vec}=(2j+1)$.  In the PBC case $j=\frac{1}{2}$  and the results correspond
to those given in Table III for the excitations relative to the ground state, 
while for APBC, $j=0, 1$ and they correspond to
those given in Table IV.  The ${\rm O}({1\over l})$ corrections to the ground state energy,
which depend on the central charges $c_{\rm sc}$ and $c_{\rm vec}$, are
also in agreement with those found in the previous section.\par
 When we put the vector and scalar parts together,
the finite-size excitation spectrum of the full model is given by
\begin{eqnarray}
&& E(j,\Delta_{\rm Ising})=E_{\rm sc}+E_{\rm vec}\nonumber \\ &&
  =\frac{\pi v_{\rm F}}{l}
   \left[-\frac{1}{12}+\frac{j(j+1)}{4}+\Delta_{\rm Ising}
         +n_{\rm sc}+n_{\rm vec}\right].
\end{eqnarray}

Not all the combinations of the 
spin and charge degrees of freedom are allowed. Only those combinations
representing composite fermions corresponding to the
$O(4)$ symmetry of our original free fermion model are allowed. 
To satisfy this condition the vector part and scalar 
part should have the same boundary conditions. 
Therefore, the possible combinations of $(j,\Delta_{\rm Ising})$ are $(0,I)$, 
$(0,\epsilon)$ or $(1,I)$, and $(1,\epsilon)$ for the APBC 
($ m\in \bf{Z}+\frac{1}{2}$), while $(\frac{1}{2},\sigma)$ for the PBC 
($ m\in \bf{Z}$). The finite-size excitation spectra of the non-interacting 
case for  PBC and APBC are shown in Table V and Table VI separately.

We now turn to the case of the interacting model $J\ne 0$. For a 
particular value of the coupling $ J=J^*=v_{\rm F}/2 $ it is possible to 
incorporate the impurity spin and define a new spin density  
$\vec{j'}(x)\equiv\vec{j}(x)+2\pi\delta(x)\vec{s}_d $
such that the impurity spin operator $\vec{s}_d$ disappears explicitly from the
Hamiltonian. The vector part of the Hamiltonian  
is now formally the same as for the non-interacting model,
$$ H_{\rm vec}^*=\frac{\pi v_{\rm F}}{4l}\sum_{m=-\infty}^{\infty}
              :\vec{j'}(-m)\cdot\vec{j'}(m): .$$ 
 The new spin-density operator $\vec{j'}(x)$ 
obeys the same $SU(2)_2$ commutation 
relations as the old one $\vec{j}(x)$ so that at the special point $J=J^*$ 
the same spectrum of excitations are generated  for $H_{\rm sc}+H_{\rm vec}^*$ 
in the conformal field algebra
as in the non-interacting case. The value of the coupling $J^*$ is very large, 
being of the order of the band width of the conduction electrons, so it corresponds to a 
strong coupling limit of the original model (see Fig.\ 4). We are interested
in the weak coupling limit of the original model. However, it might describe
the fixed point Hamiltonian in a Wilson type of
renormalization group calculation (where the energy scale is progressively reduced
by eliminating the higher energy states), similar to the case of the s-d (Kondo) model
where the fixed point is the  strong coupling one. 
This was conjectured in earlier work \cite{cit}, and  would be consistent with
the expectation that the fixed point Hamiltonian should be conformally invariant.
In the next section we look at the
results of explicit numerical renormalization  group calculations that confirm this. 
Before doing so, however, we will look at physically relevant excitations
of the model at $J=J^*$. These differ from those already considered for the non-interacting
system because, when the impurity spin has been absorbed, the allowed combinations of
the vector and scalar excitations are not the same. When we change 
from $\vec{j}(x)$ to
$\vec{j^{\prime}}(x)$, an extra spin ${1\over 2}$ degree of freedom is added to the 
$SU(2)_2$ spin density whereas the quantum numbers of the scalar part are unchanged.
The changes in the vector part 
follow the $SU(2)_2$ fusion rule \cite{al}:
$$ 0\rightarrow \frac{1}{2}, \hspace{0.5cm} \frac{1}{2}\rightarrow 0\oplus 1, \hspace{0.5cm}
     1\rightarrow \frac{1}{2}.$$
As a result, the boundary conditions of the vector part are changed compared to
the non-interacting cases: the APBC changes to PBC, while PBC to APBC. 
Therefore, the boundary conditions of the vector chain and scalar chain are 
no longer the same. We now have two sets of 
excitation spectra, which correspond to the combination of the vector part 
with PBC and scalar part with APBC, or vice versa.
 These two spectra, which are distinguishable,
 are displayed in the Table VII and Table VIII, 
respectively. We will relate these results to those of the numerical renormalization group
in the next section.\par

\section{Numerical Renormalization Group (NRG)}\par

The numerical renormalization group method was developed by Wilson
for the Kondo problem \cite{Wil75} and later applied to the single impurity
Anderson model by Krishnamurthy {\it et al.} \cite{Kri80}.
It is based on the mapping of the  model to a semi-infinite chain with the
spin or impurity on one end of the chain coupled to the first conduction
electron site via an exchange coupling $J$ or  a hybridization $V$.
The hopping matrix elements $t_n$ between neighbouring sites along the chain approach a
constant value for large $n$ as long as the mapping is performed exactly. In
the numerical renormalization group, however, certain irrelevant degrees of
freedom are neglected which results in an exponential decay 
$t_n \propto \Lambda^{-n/2}$. The origin of the parameter $\Lambda$ is the
logarithmic discretization of the continuous conduction band
with the energy mesh $\epsilon_n =\pm \Lambda^{-n}$, $n=0,1,2,\ldots$.
Details of the derivation of the semi-infinite
chain form used here  can be found in \cite{Wil75} and \cite{Kri80}.\par

Important for the discussion in this paper is the $\Lambda$-dependence of
the results due to the logarithmic discretization. As has been shown
earlier, the physical properties calculated from the NRG depend only weakly
on the discretization parameter $\Lambda$ 
as long as the number of states taken into account is not too
small (usually 500 - 1000).
On the other hand, the many-body energy levels at the fixed points 
 are $\Lambda$-dependent. In order to compare the fixed-point
spectra with the finite size spectra obtained from the CFT, we have to take
the limit $\Lambda\to 1$. This will be discussed in the following 
subsection.\par

\subsection{The non-Fermi liquid fixed point}

We want to concentrate on the fixed point spectra itself, not on the
deviations responsible for the $\ln T$-terms in the specific heat and the
susceptibility which will be considered in Sec.\ V.
A flow diagram (see also \cite{bh}) showing the lowest lying many-body energy levels is presented
in Fig.\ 5 for the isotropic case $V_{\rm a} = V/2$.

The value of the discretization parameter is $\Lambda = 2.5$. This diagram
shows the crossover from the free-orbital fixed point via the local-moment one to the
non-Fermi liquid fixed point. The influence of the local moment
fixed point is not great due to  the relatively small value of $U$. The excitations
of this fixed point are quite different to those found at the Fermi liquid fixed point for
$V_{\rm a} \ne V/2$.\par

In the previous section, the $\sigma$-$\tau$ model was written in terms of
Majorana fermions and it was pointed out that
in the isotropic case the scalar part of the
conduction electron chain is decoupled from the impurity,
and that the strong coupling affects only the vector part.
As $J\sim V^2/U$, this corresponds to the limit $V\to \infty$
of the Anderson model, where V is the hybridization between the vector
part of the impurity and the first conduction electron site ($V_0=0$).
If we start with a total odd number of sites (including the impurity) the
effective number of sites remaining is odd in the 
vector part and even
in the scalar part (see also Fig.\ 4).
The effective reduction in the length of the conduction chain at strong coupling
changes the excitation spectrum of the vector part in the same way as a change of 
boundary condition on the free periodic chain. 
Hence, the combined many-body excitations at this strong coupling point are 
the same as those calculated via the CFT (see Table VIII).\par
As the model used in the NRG calculations depends on $\Lambda$ through
the hopping matrix element $t_n\sim\Lambda^{-n/2}$, the single particle 
excitations for the free chain for this model depend on $\Lambda$. Our 
calculations for
the free chain in Sec.\ II and III, however, correspond to $\Lambda=1$. 
Having identified the
low energy fixed point of the $\sigma$-$\tau$ model in terms of the free 
Majorana chain (with changed boundary conditions), we can use this information 
to build up the many-body excitations for $\Lambda=2.5$, from the single 
particle ones ($\Lambda=2.5$) and compare these with those found in the 
numerical renormalization group calculations ($N\to \infty$).
In Fig.\ 6, the lowest lying many-body energies are calculated from the 
single particle ones for $\Lambda$ in the range, $1<\Lambda<4$. We can see 
 from this figure that the non-equally spaced excitations  go over
smoothly to the equally spaced spectrum as $\Lambda\to 1$. The complete 
agreement of this spectrum (calculated from the single particle spectrum
$\Lambda=2.5$) with that calculated from the full
NRG calculation confirms the identification of the low energy 
fixed point of the $\sigma$-$\tau$ model with the strong coupling limit.
(However, the NRG gives a ground state degeneracy of four in contrast to the CFT
result  $dg=2$ (see Table VIII); the origin of this discrepancy is still unclear.)

We want to point out that, in our case, in order to obtain Fig.\ 6 it was not necessary to 
run the full NRG-program in the entire $\Lambda$-range.
As soon as the relation of single particle and many-body energies has been
verified for a certain value of $\Lambda$ 
 the many-body energies can be calculated from the single particle
ones for $\Lambda=1$ to compare with the CFT.
This is in contrast to the work of Affleck {\it et al.} \cite{alpc} who 
studied the two channel Kondo
model with the NRG. In their case, no relation was made between the fixed
point spectrum and the single particle energies of free conduction electron
chains so to find the many-body energies to compare with the CFT 
they had to run  the full NRG program for $\Lambda \to 1$. 
This  causes  numerical problems
as the sequence of iterations does not converge 
in this limit. 

If we start off with a total even number of sites, we get agreement with the CFT results
with periodic boundary conditions for the vector part and 
antiperiodic boundary conditions for the scalar part (see Table VII).

\subsection{The Fermi liquid fixed point}

For any value of $V_{\rm a}\ne V/2$ the system flows to the Fermi liquid
fixed point of the standard single impurity Anderson model defined by  
$V_{\rm a}=0$ (see the discussion in \cite{bh}). Both scalar and vector
part of the first conduction electron site are now strongly coupled to the
impurity so that the number of effective sites is the same in both the remaining scalar and
vector part. The combination of scalar and
vector part with the same boundary condition describes the Fermi liquid
fixed point, as discussed in the previous section. Again, both the fixed
point spectrum and the degeneracies (in the limit $\Lambda\to 1$)
     are in agreement with the CFT results given in Tables V and VI.
\par

\section{Leading order corrections to the fixed points}\par

We are primarily interested in the behaviour of the models at the non-Fermi
liquid fixed point, which corresponds to $J_1=J_2$ for the $\sigma$-$\tau$
model and $V_{\rm a}=V/2$ for the O(3) Anderson model. 
Low order perturbation theory and the  numerical renormalization
 group calculations give  a zero point entropy of $\frac{1}{2}\ln 2$,
a logarithmic temperature dependence of the specific heat
       $C(T) \propto T\ln T$ , and a logarithmic temperature dependence of the `spin plus
      isospin' susceptibility $\chi_T (T) \propto \ln T $. In this section
we look at arguments that help us to identify the leading irrelevant operators
at this fixed point that generate this behaviour.  \par
Because the low energy fixed point of the $\sigma$-$\tau$ model is the strong
coupling one, as is the case for the one channel  s-d or Kondo model, 
we can use the analogies between these two to derive an expression for the
 leading irrelevant operator responsible for the low temperature
behaviour.  For the ordinary 
single-channel Kondo model, the leading irrelevant operator is simply  
${\vec{J}}^2(x=0)$, the squared conserved spin density operator, which is the 
only $SU(2)$ spin invariant operator with scaling dimension 2. For the present 
reduced model, we know that the leading irrelevant operator already exists in 
the non-interacting case where $H_{\rm vec}$, in terms of
three-component Majorana fermions with an 
$SU(2)_2$ Kac-Moody algebra, has a supersymmetry associated with 
the invariance under an 
exchange of the spin and isospin degrees of freedom, and the 
corresponding supersymmetric current operator can be defined by $\vec{J}_s(x)=i\sqrt{2}\vec{\psi}(x)+\theta\vec{j}(x)$, where $\theta$ is a 
 coordinate which anticommutes with the Majorana operators ($\theta^2=1$)
\cite{abkw}. Thus, we have 
${\vec{J}_s}^2(x)\sim \theta\psi_1(x)\psi_2(x)\psi_3(x)$, and from the free 
field theory we know ${\vec{J}_s}^2(x)$ is the unique O(3) invariant 
operator with scaling dimension 3/2. Therefore, we are led to identify the 
leading irrelevant operator  for the reduced 
single-channel impurity model as 
${\vec{J}_s}^2(x=0)\sim \theta\psi_1(0)\psi_2(0)\psi_3(0)$. As $\theta$ is an
anticommuting coordinate one can make a connection with the leading irrelevant
operator as calculated by Coleman {\it et al.} \cite{cit}. In the $1/J$ expansion  
they find a leading order  interaction term of the form  $\Phi(0)\psi_1(1)\psi_2(1)\psi_3(1)$ with 
$\Phi(0)=\psi_1(0)\psi_2(0)\psi_3(0)$. As $\Phi(0)$  is an anticommuting
operator it acts in the same way as $\theta$, so one can conclude that they
are essentially the same. We discuss this more fully later.\par 

The numerical renormalization group results  we have described are for the O(3) Anderson model
 which, as we discussed in the introduction, is equivalent
 to the $\sigma$-$\tau$ model in the large $U$ (local moment)
regime with $J\sim  V^2/U$. 
We have, therefore, interpreted  the strong coupling fixed point, 
in Anderson model terms, as corresponding to the limit $V\to \infty$, which
effectively removes the first conduction site from the vector Majorana fermion
chain. We end up for both models with the fixed point described by
free  vector and scalar chains of Majorana fermions with effective  lengths which  
differ by one site. However, the  NRG results for the Anderson model for $V_{\rm a}=V/2$ 
show the
same non-Fermi liquid behaviour {\em for all values 
of} $U$. It is possible to give an equivalent interpretation
 of this fixed point 
as the $U=0$ limit of the Anderson model.
This situation  is essentially the same as the $V\to\infty$ one as it corresponds
 to   free vector   and  scalar Majorana fermion
chains whose lengths differ by one site, as only the vector  Majorana chain
is coupled to the impurity. A similar situation occurs for the standard single channel
Anderson model where the fixed point can be interpretated as a Fermi liquid fixed point
for all values of $U$. We develop this approach further in Sec.\ VII.\par

We now briefly look at the arguments used in the NRG to identify the leading irrelevant operator at the 
non-Fermi liquid fixed point for the O(3) symmetric Anderson model. We take $H_N$ to be
the non-interacting model ($U=0)$ with an $N$ site conduction chain, and consider a
perturbation of the form,
\begin{equation}
   \delta H_1 = -\tilde{U} \psi_0(-1)\psi_1(-1)\psi_2(-1)\psi_3(-1)
   \label{eq:def_delta}
\end{equation}
where $\psi_0(-1)$ is an additional Majorana-fermion that does not couple
to the scalar chain starting at site 0 (see Fig.\ 7), so our effective impurity
site here corresponds to $N=-1$. Wilson devised a way of determining the degree of
irrelevancy (or relevancy) of a particular operator by considering how the operator scales 
on increasing the length of the chain $N\to N+2$, when expressed in terms of the creation
and annihilation operators for diagonalized states of $H_N$ and $H_{N+2}$, respectively.
Because of the overall scaling factor the appropriate terms to consider are
$\delta H^\prime = \Lambda^{\frac{N-1}{2}}\delta H_1(N) $  and $\delta H^{\prime\prime} = 
\Lambda^{\frac{N+1}{2}}\delta H_1(N+2) $. Following this line of reasoning we find

\begin{equation}
  \delta H_1^{\prime\prime} = \Lambda^{-1/2} \delta H^\prime_1,
\end{equation}
so that the effective perturbation on the fixed-point Hamiltonian
is reduced by a factor $\Lambda^{-1/2}$ when $N$ is increased by 2.
This result is in contrast to the standard SAM where a perturbation of the
form (\ref{eq:def_delta}) decreases by a factor of $\Lambda^{-1}$.
This difference is due to the coupling of {\it all} four components
of the Majorana fermions to the rest of the chain.
Another difference  is that 
(\ref{eq:def_delta}) is the only leading irrelevant perturbation
to the non-Fermi liquid fixed point whereas
in the standard case a perturbation in form of a hybridization between
site -1 and site 0 also reduces with $\Lambda^{-1}$.

In the O(3) symmetric Anderson model the hybridization term that couples the vector parts of site
-1 and 0 has the form
\begin{equation}
   \delta H_2 = i\tilde{V}\sum_{\alpha=1}^3
\psi_\alpha(-1)\psi_\alpha(0)
   \label{eq:def_delta2}.
\end{equation}
Using the same type of argument as just given we can show that
the corresponding scaled quantities,  $\delta H_2^{\prime}$,
$\delta H_2^{\prime\prime}$, for the $N$ and $N+2$ site models 
are related via
\begin{equation}
  \delta H_2^{\prime\prime} = \Lambda^{-1} \delta H^\prime_2 \ .
\end{equation}
The perturbation $\delta H_2$ therefore is more irrelevant
than $\delta H_1$.
We know from general arguments advanced by Wilson that other
local operators involving more sites, or more complicated operators
at the initial sites, only give terms with a higher degree of irrelevancy.\par

In the definition of $\delta H_2$ we did not take into account the coupling
between the scalar parts of site -1 and 0 
\begin{equation}
   \delta H_3 = i\tilde{V}_0
\psi_0(-1)\psi_0(0)
   \label{eq:def_delta3} \ .
\end{equation}
This perturbation scales as 
$\delta H_3^{\prime\prime} = \Lambda^{1/2} \delta H^\prime_3$
and drives the system away from the non-Fermi liquid fixed point.
However, it only exists in the anisotropic case $V_{\rm a}\ne V/2$
where we know from the NRG results that the non-Fermi liquid fixed point
is unstable.\par
Here we have regarded the fixed point to correspond to the $U=0$ model. 
If we had considered it to correspond to $V\to\infty$, nothing would have essentially
changed in the argument except the way we label the initial sites in the chains.\par
Putting together these leading irrelevant terms with the fixed point Hamiltonian
gives us a renormalized form of the O(3) Anderson model. We postpone discussion of 
how to parameterize this model, and how to calculate the low temperature thermodynamics
about the fixed point to Sec.\ VII. Before that we consider some results for the O(3)
model is its "bare" form as defined in Eq.\ (\ref{hamm}).\par

\section{ Perturbation Theory and Ward Identities}\par

In this section we consider the perturbation theory in powers of $U$ for the O(3)
Anderson model. In an earlier paper \cite{zh} it was shown that there are diagrams for
the model with $V_0=0$ that give vertex corrections that diverge as ${\rm ln}(T)$ as $T\to 0$,
indicating that the $U$ term behaves like a marginally relevant operator. Here we show that
there are other vertex corrections of the same order which also give ${\rm ln}(T)$ terms,
such that the net effect is that the ${\rm ln}(T)$ terms cancel. This alters our conclusions
about the stability of the fixed point in the presence of $U$. We show that similar terms
that lead to higher powers of ${\rm ln}(T)$  for the susceptibility $\chi_j$ also cancel.
We also derive a Ward identity that allows us to demonstrate that the Wilson ratio
for $\chi_j/\gamma$ is ${8\over 3}$ independent of $U$, in agreement with the NRG results
 \cite{bh}. \par
The most direct way to exploit the symmmetries of the model,
and deduce a  Ward identity, is to use the functional integral approach, which can also be 
used to generate the perturbation theory. 
We can make the standard transformations and express the partition
 function $Z$ for the model as a functional integral over Grassmann variables associated
with both the impurity and conduction electron Majorana states,
\be 
Z=\int \prod_{\alpha=0}^3 D[d_{\alpha}] \prod_{n=0}^{\infty} D[\psi_{n,\alpha}]e^{-S},
\ee
where  the action $S$ is given by
\bea
 S&=&\int_0^{\beta}d\tau\Bigg[{1\over 2}\sum_{\alpha=0}^3 \big\{ d_{\alpha}(\tau)
\partial_{\tau}d_{\alpha}(\tau) 
 \nonumber \\ 
 & & \hspace{2cm}+\sum_{n=0}^{\infty}\psi_{\alpha}(n,\tau)
\partial_{\tau}\psi_{\alpha}(n,\tau)\big\}\nonumber \\
& &\hspace{1cm}+\sum_{\alpha=0}^3\sum_{n=0}^{\infty}t_n
\psi_{\alpha}(n+1,\tau)\psi_{\alpha}(n,\tau)\nonumber \\
 & &\hspace{1cm}+iV\sum_{\alpha=1}^3 \psi_{\alpha}(0,\tau)d_{\alpha}(\tau)+iV_0\psi_{0}(0,\tau)d_{0}(\tau)
\nonumber \\
& &\hspace{1cm}+U d_{1}(\tau)d_{2}(\tau)d_{3}(\tau)d_{0}(\tau) \Bigg].\label{pf}
\eea
Because the action is purely bilinear in the conduction electron Grassmann variables 
these can be integrated over to give a reduced action  $S_{\rm red}$  which is expressed 
in terms of the impurity Grassmann variables only, 
\bea & & S_{\rm red}=\nonumber \\
& & -{1\over 2}\int_0^{\beta}d\tau\int_0^{\beta}d\tau^{\prime}
\sum_{\alpha=0}^3 d_{\alpha}(\tau)[\partial_{\tau}\delta(\tau-\tau^{\prime})
+R_{\alpha}
(\tau-\tau^{\prime})]d_{\alpha}(\tau^{\prime}) \nonumber \\
 & & +U\int_0^{\beta}d\tau d_{1}(\tau)d_{2}(\tau)d_{3}(\tau)d_{0}(\tau). \label{fia}
\eea
This expression differs from that for an isolated impurity only  through the term
 $R_{\alpha}(\tau-\tau^{\prime})$.  The form for $R_{\alpha}(\tau-\tau^{\prime})$  
can most conveniently be expressed in terms of its Fourier coefficient $R_{\alpha}(\omega_n)$. 
The functional integral can be re-expressed as an integration over  the Fourier coefficients of 
the Grassmann variables $d_{\alpha}(\omega_n)$, where $d_{\alpha}(\tau)=
{1/ \beta}\sum_n d_{\alpha}(\omega_n)e^{-i\omega_n\tau}$ with $\omega_n=(2n+1)\pi/\beta$ so as 
to satisfy antiperiodic boundary conditions
$d_{\alpha}(\beta)=-d_{\alpha}(0)$. The Fourier series for $R_{\alpha}(\tau-\tau^{\prime})$ is 
$R_{\alpha}(\tau-\tau^{\prime})={1/\beta}\sum_n R_{\alpha}(\omega_n)e^{-i\omega_n(\tau-\tau^{\prime})}$. For a 
conduction band $\rho_0(\epsilon)$ with $-D <\epsilon<D$, $R_{\alpha}(\omega_n)$ is given by
\be R_{\alpha}(\omega_n)={1\over \pi}\int_{-D}^{D}{{\Delta_{\alpha}(\epsilon)d\epsilon}
\over{i\omega_n+\mu-\epsilon}}
 -i\Delta_{\alpha}(i\omega_n+\mu){\rm sgn}(\omega_n)\label{Rdel}\ee
where $\Delta_{\alpha}(\epsilon)=\pi V_{\alpha}^2\rho_0(\epsilon)$. In the flat wide band limit,
 $\rho_0(\epsilon)$
is independent of $\epsilon$ and $D\to \infty$, $\Delta_{\alpha}(\epsilon)$ becomes a 
constant.\par
In the  O(3) model  we have $V_{1}=V_{2}=V_{3}=V$ so 
 we denote $\Delta_{\alpha}=\Delta$ for $\alpha=1,2,3$.
 When $V_0=0$ in Eq.\ (\ref{hamm})  the localized d,  ${\alpha=0}$ Majorana 
fermion is decoupled from the conduction electrons and $\Delta_{0}=0$.
In the functional integral  (\ref{fia})  the $\alpha=0$ Majorana fermion  then has a zero energy 
mode at $T=0$ and scattering 
of  the other Majorana fermions with this zero energy excitation  leads to
infrared singularities and  to non-Fermi liquid behaviour \cite{cit,zh}. \par

\subsection{Low order perturbation theory}\par

We can develop the perturbation theory in $U$ of  reference \cite{zh} 
for  the impurity Green's functions in the functional integral
formalism from the generating function $Z(\eta_{\alpha}(\tau))$.  
This is constructed by adding a coupling to a local Majorana Grassmann field
 $\eta_{\alpha}(\tau)$ to the reduced action  (\ref{fia})  of the form,
\be -\int_0^{\beta}d\tau\sum_{\alpha=0}^3\eta_\alpha(\tau)d_{\alpha}(\tau).\ee
The Green's functions are then obtained by taking the appropriate functional derivatives
with respect to the fields.  The thermal Green's function
 $G^{(0)}_{\alpha\beta}(\tau-\tau^{\prime})$  for the impurity 
 Majorana fermions for $U=0$ is given by
\bea 
G^{(0)}_{\alpha\beta}(\tau-\tau^{\prime})&=&-\langle T_{\tau} 
d_{\alpha}(\tau)d_{\beta}(\tau^{\prime})\rangle_0 \nonumber \\&=&
{1 \over \beta} \sum_n G^{(0)}_{\alpha\beta}(\omega_n)e^{-i\omega_n(\tau -\tau^{\prime})},
\eea
where 
\be G^{(0)}_{\alpha\beta}(\omega_n)={\delta_{\alpha\beta}\over {i\omega_n+
i\Delta_{\alpha}{\rm sgn(i\omega_n)}}},\label{gf}\ee
with $\omega_n=(2n+1)\pi/\beta$ in the wide band limit.\par
We review the earlier results for the four vertex 
$ \Gamma_{0,1,2,3}(\omega_{n_0},\omega_{n_1},\omega_{n_2},\omega_{n_3})$,
which analytically continued to real frequencies  and then evaluated at zero frequency.
 Diagrams for this vertex up to third order are shown in Fig.\ 8.
In third order of $U$, the interaction vertex corrections are given by
\begin{eqnarray}
  \Gamma^{(3)}_{0,1,2,3}(0,0,0,0)
  =&& -3\frac{U^3}{\beta^2} \sum_{\omega_1,\omega_2}
      G_{\alpha}^2(\omega_1) G_{\alpha}(\omega_2) G_0(\omega_2)
  \nonumber \\ && \hspace{-3cm}
+3\frac{U^3}{\beta^2}\lim_{\omega\rightarrow 0}\sum_{\omega_1,\omega_2}
      G_{\alpha}(\omega_1) G_{\alpha}(\omega_2) G_0(\omega_2)
          G_{\alpha}(\omega_1+\omega_2-\omega)
 \nonumber \\&&  \hspace{-3cm}
+3\frac{U^3}{\beta^2}\lim_{\omega\rightarrow 0}\sum_{\omega_1,\omega_2}
       G_{\alpha}(\omega_1)G_{\alpha}(\omega_2)^2G_0(\omega_1+\omega_2-\omega),
\end{eqnarray}
which correspond to the Feymann diagrams (b), (c), (d) in Fig.\ 8, respectively,
where the Green's function propagators  are all for $U=0$, as given in Eq.\ (\ref{gf}).
 Diagrams (b) and (c) can be severed into 
separate diagrams by cutting a pair of lines, one of which is the propagator 
of the $\alpha=0$ Majorana fermions. Such diagrams give singular 
contributions to the interaction vertex  (parquet diagrams). 
They can be evaluated as follows
\begin{eqnarray}
 I_b 
 =&& -3 U^3\left [-\frac{1}{\beta}\sum_{\omega_n}G^2(\omega_n) \right ]
  \left [-\frac{1}{\beta}\sum_{\omega_n}G(\omega_n)G_0(\omega_n) \right]
 \nonumber \\ &&
 \approx -3U\left(\frac{U}{\pi\Delta}\right)^2
      \left[\Psi(\frac{1}{2}+\frac{\Delta}{2\pi T})-\Psi(\frac{1}{2})\right]
 \nonumber \\ &&
 \approx -3U\left(\frac{U}{\pi\Delta}\right)^2
\ln\left(\frac{\Delta}{T}\right),
    \hspace{1cm} {\rm for} \hspace{.5cm} T\ll\Delta.
\end{eqnarray}
While for the second term, we have to complete the summations over the 
internal frequencies, and then let the external frequency go to zero. After 
that, we obtain the following triple integral
\begin{eqnarray}
  I_c
 &=&-\frac{3U^3}{4}\int_{-\infty}^{\infty}d\epsilon_1 d\epsilon_2 d\epsilon_3
  \rho(\epsilon_1)\rho(\epsilon_2)\rho(\epsilon_3) \nonumber \\
 &\times &  \frac{1}{\epsilon_2 (\epsilon_1+\epsilon_2-\epsilon_3)}
  \frac{\cosh(\frac{\epsilon_1+\epsilon_2-\epsilon_3}{2T})}
       {\cosh(\frac{\epsilon_1}{2T})\cosh(\frac{\epsilon_2}{2T})
         \cosh(\frac{\epsilon_3}{2T})}.
\end{eqnarray}
By changing the variables, we can reduce the integral, and
then extract the leading singular contribution, which is
\begin{eqnarray}
 I_c =&&  3U(\frac{U}{\pi\Delta})^2\int_0^{\Delta}d\epsilon_1 
          \frac{\tanh(\epsilon_1/2T)}{\epsilon_1} \nonumber \\ &&
    \approx 3U(\frac{U}{\pi\Delta})^2\ln(\frac{\Delta}{T}),
\end{eqnarray}
where a high energy cutoff factor $\Delta$ has been introduced when calculating
the integral. \par

The above results have clearly shown that the logarithmic contributions to the 
interaction vertex corrections in the third order perturbation theory cancel 
exactly. We are left  diagram (c) in Fig.\ 8 which contributes a regular term only. 
Therefore, up to third order in $U$, there are no singular vertex corrections.     
In a renormalized perturbation theory  this vertex is multiplied by a wavefunction
renormalization factor which can be shown does have ${\rm ln}(T)$ terms.
However, the sign of this term is such that in the renormalization group equations
of the form given in \cite{zh} the renormalized interaction decreases rather than
increases so the fixed is stable in the presence of $U$ and not unstable as concluded earlier.
\par
In a similar way we can show that 
the  $({\rm ln}(T))^2$ contributions to the total spin susceptibility $\chi_j$ 
that occur to order $U^4$ also cancel. This susceptibility 
 is  the response of the impurity  to a field $H_{12}$ ($=g_j\mu_{\rm B}H$) coupled to the $j_z$ 
component of the impurity states. 
This coupling corresponds 
to an extra term in the action of the form,
\be {{iH_{12}}\over{\beta}}\int_0^{\beta}d_1(\tau)d_2(\tau)d\tau ,\label{cj}\ee
and $\chi_j$ is given by
\be \chi_j=-i(g_j\mu_{\rm B})^2{{\partial \langle d_1d_2\rangle}\over {\partial H_{12}}}.\ee
The first singular contribution to $\chi_j$ comes from 
the second order in $U$, described by the diagram (a) in Fig.\ 9. It is 
straightforward to evaluate it, and obtain
\begin{equation}
 {{\chi_j^{(2)}}\over {(g_j\mu_{\rm B})^2}}=(\frac{U}{\pi\Delta})^2\frac{1}{\pi\Delta}\ln(\frac{\Delta}{T}).
\end{equation}
When we consider the fourth order in $U$, higher logarithmic terms appear, and
they are described by the diagrams (b) and (c) in Fig.\ 9. According to the 
previous argument, these two diagrams are also parquet ones, which will produce
the singular contributions to $\chi_j$. 
The evaluation of the diagram (b) is easier, and we can get
\begin{eqnarray}
 I'_b=&& U^4\left[\frac{1}{\beta}\sum_{\omega_1}G^2_{\alpha}(\omega_1)\right]^3
  \left[\frac{1}{\beta}\sum_{\omega_2}G_{\alpha}(\omega_2)G_0(\omega_2)\right]^2 \nonumber \\ &&
 \approx -(\frac{U}{\pi\Delta})^4\frac{1}{\pi\Delta}\ln^2(\frac{\Delta}{T}).
\end{eqnarray}
While for the diagram (c), it is more complicated to calculate its 
contribution. It is denoted as the following term
\begin{eqnarray}
 I'_c&=&U^4(\frac{1}{\pi\Delta})^2(\frac{1}{\beta})^3
    \sum_{\omega_1,\omega_2,\omega_3}G_{\alpha}(\omega_1)G_0(\omega_1)
  G_{\alpha}(\omega_2)\nonumber \\ 
     &\times& \hspace{0.5cm} G_0(\omega_2)G_{\alpha}(\omega_3)
     G_0(\omega_2-\omega_1-\omega_3).
\end{eqnarray}
Here we also have to finish the summmations over frequencies first, and then 
we find that
\begin{eqnarray}
 I'_c&=&\frac{U^4}{8}(\frac{1}{\pi\Delta})^2
\int_{-\infty}^{\infty}d\epsilon_1 d\epsilon_2 d\epsilon_3 d\epsilon_4
  \rho(\epsilon_1)\rho(\epsilon_2)\rho(\epsilon_3)\rho(\epsilon_4)
\nonumber \\
 &\times&  \frac{1}{\epsilon_3\epsilon_4 (\epsilon_1+\epsilon_2-\epsilon_3+\epsilon_4)}
  \nonumber \\
 &\times&  \frac{\sinh(\frac{\epsilon_1+\epsilon_2-\epsilon_3+\epsilon_4}{2T})}
       {\cosh(\frac{\epsilon_1}{2T})\cosh(\frac{\epsilon_2}{2T})
         \cosh(\frac{\epsilon_3}{2T})\cosh(\frac{\epsilon_4}{2T})}.
\end{eqnarray}
After that, by changing variables we can reduce the integral further and
identify the leading singular contribution, which is
\begin{eqnarray}
 I'_c &&= (\frac{U}{\pi\Delta})^4\frac{1}{\pi\Delta} 
    \int_0^{\Delta}d\epsilon_1\frac{\tanh(\epsilon_1/2T)}{\epsilon_1}
     \int_0^{\Delta}d\epsilon_2\frac{\tanh(\epsilon_2/2T)}{\epsilon_2} 
 \nonumber \\ &&
  \approx (\frac{U}{\pi\Delta})^4\frac{1}{\pi\Delta}\ln^2(\frac{\Delta}{T}).
\end{eqnarray}
Therefore, the squared logarithmic contribution terms in the fourth order 
perturbation for $\chi_j$ cancel exactly, but there are still the 
logarithmic terms in this order. \par
\subsection{Ward Identity}\par
 
To derive the Ward identity we exploit the symmetry noted earlier that 
the Hamiltonian is invariant under an O(3) orthogonal transformation in
the space spanned by the 1,2,3 Majorana fermions. We first of all look at the effect of using
such a transformation to change the variables in the action. We look at a particular one parameter
transformation, corresponding to a
rotation through an angle $\theta$ in the 1,2 plane about axis 3,
\bea
 d_1(\tau)&=&{\rm cos}\theta(\tau) 
  d_1^{\prime}(\tau)-{\rm sin}\theta(\tau) d_2^{\prime}(\tau),\nonumber \\
d_2(\tau)&=&{\rm sin}\theta(\tau) d_1^{\prime}(\tau)+{\rm cos}\theta(\tau) 
d_2^{\prime}(\tau) \eea
For $\theta$ independent of $\tau$, and in the absence of external source fields, the
partition function is invariant under this change of variables. However, when $\theta$ 
is taken to be a local $\tau$ dependent transformation a new term in the action is generated,
and there are further new terms arising from the couplings to the external source fields, as
these are also not invariant under the transformation. Apart from these extra terms the expression
for the generating function has the same form in terms of the new variables (and the same measure)
so these extra contributions must cancel. The equation  for the cancellation of these two
terms leads to the Ward identity. We work to first order only in $\theta$. The extra term 
to the partition function from the coupling to the external source fields 
can be written in the form,
\be \theta(\tau)(\eta_{1}(\tau)\langle d_2(\tau)\rangle-\eta_2(\tau) \langle d_1(\tau)\rangle)\ee
where we have divided by $Z$ in order to express the result in the form of an expectation value
(with respect to the action for $\theta=0$). In terms of the Fourier coefficients this becomes
\bea 
{1\over{\beta}^2}\sum_{n,m}\theta(\Omega_m)\Big(\eta_1(-\omega_n)\langle d_1(\omega_n-\Omega_m)
\rangle-\nonumber \\
\eta_2(-\omega_n) \langle d_2(\omega_n-\Omega_m)\rangle\Big)
\eea
 $\theta(\Omega_m)$ is the Fourier coefficient in the expansion of $\theta(\tau)$
\be \theta(\tau)={1\over \beta}\sum_m \theta(\Omega_m)e^{-i\Omega_m\tau},\ee
with $\Omega_m=2m\pi/\beta$ and $ m$ an integer, as $\theta(\tau)$ is required to satisfy 
the periodic
boundary condition $\theta(\beta)=\theta(0)$. 
The interaction part of the action is
invariant under the transformation, even when $\theta$ depends on $\tau$, so the other
extra term generated arises purely from the bilinear term. This extra contribution can
be written in the form,
\be {1\over{\beta}^2}\sum_{n,m}\theta(\Omega_m)\langle d_1(-\omega_n)
 d_2(\omega_n-\Omega_m)\rangle)F_{12}(\omega_n, \Omega_m),\ee
where $F_{12}(\omega_n, \Omega_m)$ is defined by
\be F_{\alpha\beta}(\omega_n, \Omega_m)=\Gamma_{\alpha\alpha}^{(2,0)}(\omega_n)-
\Gamma_{\beta\beta}^{(2,0)}(\omega_n-\Omega_m)\ee
with $\Gamma_{\alpha,\beta}^{(2,0)}(\omega_n)=[G_{\alpha,\beta}^{(0)}(\omega_n)]^{-1}$.
We now equate the sum of these two terms to zero. We can use the fact that the resulting
 equation holds for arbitrary $\theta(\Omega_m)$, so the coefficient of each component
must vanish. This leads to the equation,
$$ \sum_{n'}(\langle d_1(-\omega_n')
 d_2(\omega_n'-\Omega_m)\rangle)F_{12}(\omega_n', \Omega_m)$$
\be =\sum_{n'}(\eta_2(\omega_n'-\Omega_m)\langle d_1(-\omega_n'))
\rangle-\eta_1(-\omega_n') \langle d_2(\omega_n'-\Omega_m)\rangle)\ee
which is the basic Ward identity. The terms in this equation all vanish in the 
absence of the external source terms so to derive any useful equations from this
for our original system without sources we must  functionally differentiate
with respect to $\eta_1(-\omega_n)$ and  $\eta_2(\omega_n-\Omega_m)$.
On carrying out this differentiation and then equating
the external source terms to zero we obtain the equation,
\bea
& &-(\langle d_1(-\omega_n)d_1(\omega_n)
\rangle- \langle d_2(-\omega_n+\Omega_m)d_2(\omega_n-\Omega_m)\rangle)
\nonumber \\ &=&
 {1\over{\beta}}\sum_{n'}\langle d_1(\omega_n)
 d_2(-\omega_n+\Omega_m)d_1(-\omega_n')
 d_2(\omega_n'-\Omega_m)\rangle) \nonumber \\
& &\hspace{1cm}\times F_{12}(\omega_n', \Omega_m)
\eea
which is an expression which relates the one and two particle Green's functions.
It can be checked in the case $U=0$, where it gives a trivial identity.
For $U\ne 0$ we can express the two particle Green's function in terms of the
one particle irreducible four vertex,
\bea 
& &\Sigma_{11}(\omega_n)- \Sigma_{22}(\omega_n-\Omega_m)= \nonumber \\
&  -&{1\over{\beta}}\sum_{n'}
 \Gamma^{(4)}_{1212}(\omega_n,\omega_n-\Omega_m,\omega_n',
\omega_n'-\Omega_m)\nonumber \\
&\times & F_{12}(\omega_n', \Omega_m)G_{11}(\omega_n')G_{22}
(\omega_n'-\Omega_m)\label{Oderiv}
\eea
which is now a non trivial identity relating the self-energy of the one
particle Green's function to the irreducible four vertex.\par

We can obtain another similar identity when the coupling of the impurity to an external field
of the form (\ref{cj}) is taken into account.
Such an interaction induces an off diagonal term $\Sigma_{12}(\omega_n)$
in the self-energy. To relate these quantities we have to use Dyson's equation
which is now in matrix form,
\be [{\bf \Gamma}^{(2,0)}-{\bf \Sigma}]{\bf G}={\bf I}.\ee
The change in the Green's function $\delta{\bf G}$ to first order in the applied
field is given by
\be \delta{\bf G}={\bf G}[{\bf \Gamma}^{(2,0)}\delta{\bf G}^{(0)}{\bf \Gamma}^{(2,0)}
+\delta{\bf \Sigma}]{\bf G},\ee
and as the matrices are diagonal in the absence of the field this can be
simplified to
\be  G_{12}=G_{11}G_{22}(\Gamma^{(2,0)} \delta G_{12}^{(0)}\Gamma^{(2,0)}+\Sigma_{12}).\ee
The first term involves $\delta G_{12}^{(0)}$ the off diagonal Green's function calculated
to first order in $H_{12}$ but with the interaction term set to zero.
This is easy to calculate, and on substituting the result in the above we
find
\be  G_{12}(\omega_n)=G_{11}(\omega_n)G_{22}(\omega_n)(iH_{12}+\Sigma_{12}(\omega_n))\ee
An alternative expression for $G_{12}(\omega_n)$ to first order in $H_{12}$ 
can be deduced from the
generating function, by taking  the appropriate functional derivatives to generate
$G_{12}(\omega_n)$ and then taking a derivative with respect to the external field
$H_{12}$. We  obtain the equation,
\be G_{12}(\tau_1,\tau_2)={{iH_{12}}\over{\beta}}\int_0^{\beta}\langle T_{\tau}d_1(\tau_1)
d_2(\tau_2)d_1(\tau)d_2(\tau)\rangle d\tau .\ee
The expectation value on the right hand side of this equation is a two particle
Green's function and can be rewritten in terms of the irreducible four
vertex $\Gamma_{1212}$,
\bea  G_{12}(i\omega_n)&=&
 G_{11}(i\omega_n)G_{22}(i\omega_n) \Big(iH_{12}+{{iH_{12}}\over \beta}
\nonumber \\
&& \hspace{-1.5cm} \times
\sum_{n'}\Gamma_{1212}
(\omega_n,\omega_n,\omega_n',\omega_n')G_{11}(\omega_n')G_{22}(\omega_n')
\Big).
\eea
Equating this to our previous result and taking the limit $H_{12}\to 0$ we find
\be {{\partial \Sigma_{12}(\omega_n)}\over {i\partial H_{12}}}={1\over \beta}\sum_{n'}
\Gamma_{1212}
(\omega_n,\omega_n,\omega_n',\omega_n')G_{11}(\omega_n')G_{22}(\omega_n').\label{Hderiv}\ee
As $T\to 0$ the discrete frequencies $\omega_n$ can be replaced by a continuous variable
$\omega$ and the summations can be replaced integrals so Eq.\ (\ref{Hderiv}) becomes
\be {{\partial \Sigma_{12}(\omega)}\over {i\partial H_{12}}}=\int\Gamma_{1212}
(\omega,\omega,\omega',\omega')G_{11}(\omega')G_{22}(\omega'){{d\omega'}\over{2\pi}}.\label{hcont}\ee
We can take the same limit of Eq.\ (\ref{Oderiv}), and if we also divide by $\Omega$
and take the limit $\Omega\to 0$, and expression similar to (\ref{hcont}) will be obtained
for the derivative of the self-energy with respect to frequency (we can use the fact that
$\Sigma_{11}(\omega)=\Sigma_{22}(\omega)$). We have to be careful in taking the limit
$\Omega\to 0$ because in this limit there is a delta function contribution to the integrand
at $\omega=0$. This arises from the derivative of the ${\rm sgn}$ functions that
originate from the imaginary part in Eq.\ (\ref{Rdel}). The term in the integrand
of (\ref{Oderiv}),
\be
 F_{12}(\omega', \Omega)G_{11}(\omega')G_{22}
(\omega'-\Omega),
\ee
can be rewritten in the form,
\bea
 &-&{{[G_{11}(\omega')-G_{22}(\omega'-\Omega)]}\over \Omega}
\nonumber \\
&+&{{[\Sigma_{11}
(\omega')-\Sigma_{22}(\omega'-\Omega)]}
\over \Omega}G_{11}(\omega'-\Omega)G_{22}(\omega'),\label{above}
\eea
The first term in (\ref{above}) gives the delta function term when we take the limit
$\Omega\to 0$.  In this limit (\ref{above}) becomes
\be  G_{11}(\omega')G_{22}(\omega')+\pi(\rho_{11}(0)+\rho_{22}(0))\delta(\omega'),\ee
where $\rho_{\alpha\alpha}(0)$ is the spectral density of the  $G_{\alpha\alpha}$
Green's function at the Fermi 
level. Here we have used the fact that the Green's functions $G_{11}$ and $G_{22}$
are equal and the corresponding self-energies are equal but we write
the result in a symmetrical way. We have also taken the flat infinitely
wide conduction band limit for simplicity. The final expression for the derivative
of the self-energy
is
$$ {{\partial \Sigma_{11}(\omega)}\over {i\partial\omega}}+
{{\partial \Sigma_{22}(\omega)}\over {i\partial\omega}}=\pi(\rho_{11}(0)+\rho_{22}(0))
\Gamma_{1212}
(\omega,\omega,0,0)$$ 
\be +2\int\Gamma_{1212}
(\omega,\omega,\omega',\omega')G_{11}(\omega')G_{22}(\omega'){{d\omega'}\over{2\pi}},\label{w1}\ee
and hence we find on using (\ref{hcont}) the identity,
\bea
{{\partial \Sigma_{11}(\omega)}\over {i\partial\omega}}+
{{\partial \Sigma_{22}(\omega)}\over {i\partial\omega}}=
2{{\partial \Sigma_{12}(\omega)}\over {i\partial H_{12}}}
\nonumber \\
+ \pi(\rho_{11}(0)+\rho_{22}(0))
\Gamma_{1212}
(\omega,\omega,0,0) \label{w2}
\eea
These two equations can be used to derive an exact expression for the impurity 
contribution to the susceptibility $\chi_j$. We calculate $\chi_j$ using
\bea
{{\chi_j}\over{(g_j\mu_{\rm B})^2}}&=&-i{{\partial \langle d_1d_2\rangle}\over {\partial H_{12}}}=
\nonumber \\
& & \hspace{-1cm} -\int_{-\infty}^{\infty}G_{11}(\omega)G_{22}(\omega)\left(1
-{{\partial \Sigma_{12}(\omega)}\over {i\partial H_{12}}}\right){{d\omega}\over{2\pi}}.
\eea
We can now use (\ref{w2}) and substitute the expression for the derivative of the off diagonal
self-energy,
\bea
{{\chi_j}\over{(g_j\mu_{\rm B})^2}}=\left\{-\int_{-\infty}^{\infty}G^2_{11}(\omega)\left(1
-{{\partial \Sigma_{11}(\omega)}\over {i\partial \omega}}\right){{d\omega}\over 2\pi}
\right. \nonumber \\ 
\left.-
\rho_{11}(0)\int_{-\infty}^{\infty} G^2_{11}(\omega)\Gamma_{1212}
(\omega,\omega,0,0){{d\omega}\over{2\pi}}\right\}
\eea
where we have used the fact the Green's functions and self-energies for the vector channels
$\alpha=1,2$ are equal. The first integral is easy to evaluate because it is the derivative of 
$G_{11}(\omega)$ ($\omega \ne 0$), while for the second integral we can use (\ref{w1}) for $\omega=0$,
plus the fact that $\Gamma_{1212}(0,0,0,0)=0$  due to antisymmetry. We then obtain
the result,
\be\chi_j=(g_j\mu_{\rm B})^2
\rho_{11}(0)\left\{1-{{\partial \Sigma_{11}(\omega)}\over {i\partial\omega}}
\right\}_{\omega=0},\ee
which is an exact result for this susceptibility in the limit $T\to 0$.\par
We can obtain a similar exact result for the specific heat coefficient $\gamma$ by applying
the same line of reasoning as is used for the SAM but in terms of the Majorana Green's functions.
The result for the impurity contribution to $\gamma$ is
\be \gamma={{2\pi^2k_{\rm B}^2}\over 3} \sum_{\alpha}\rho_{\alpha\alpha}(0)\left\{1-
{{\partial \Sigma_{\alpha\alpha}(\omega)}\over {i\partial\omega}}\right\}_{\omega=0},\ee
where the sum runs over $\alpha=0,1,2,3$ for $V_{\rm a}\ne V/2$. For $V_{\rm a}=V/2$
the sum runs only over the vector components $\alpha=1,2,3$ as the scalar contribution
to the free energy gives a term in $T$ rather than $T^2$, which contributes to the residual
entropy term rather than the specific heat coefficient.\par
From these results we can derive a Wilson ratio ($R$) of $\chi_j/\gamma$ which is universal
in the case $V_{\rm a}=0$ and $V_{\rm a}= V/2$. The first case corresponds to the O(4)
symmetric model in which each Majorana fermion term makes an equal contribution so that
\be R={{4\pi^2k_{\rm B}^2}\over {3(g_j\mu_{\rm B})^2}}{{\chi_j}\over {\gamma}}=2.\label{wr1}\ee
This result is independent of $U$ in agreement with the NRG results. We also know that
in the large $U$ limit $\chi_j\to \chi_s$ where $\chi_s$ is the spin susceptibility,
as the isospin fluctuations associated with the impurity are suppressed, so this result
agrees with the  Wilson ratio of the spin susceptibility to the specific
heat coefficient $\chi_s/\gamma$ in the large $U$ limit. For $V_{\rm a}=0$ we
can derive similar exact expression for the spin and charge susceptibilities from
Ward identities \cite{hew3}, because both these quantities are conserved when the
anomalous hybridization term vanishes. We cannot do so for $V_{\rm a}\ne 0$ as
these quantities are no longer conserved. It is possible to derive some
general Ward identities for the derivatives of the self-energies of all  four
Majorana fermions when $V_{\rm a}\ne 0$ because the non-conserving
terms do not involve the two-body interaction term \cite{hew3}. However, expressions for
the spin and charge susceptibilities cannot be deduced in terms to the self-energies
evaluated at zero frequency, as they do not involve conserved quantities, so that
the information given by the Ward identities cannot be exploited in the same  way.
\par
For $V_{\rm a}= V/2$ we again find a universal Wilson ratio,
\be R= {{4\pi^2k_{\rm B}^2}\over {3(g_j\mu_{\rm B})^2}}{{\chi_j}\over {\gamma}}={8\over 3},\label{wr2}\ee
The value has changed from the SAM value simply because one of the Majorana fermions does not
contribute to the specific heat coefficient and so the results is ${8\over 3}$ rather than 
${8\over 4}$. The corresponding Wilson ratio for the spin susceptibility should have
the same value in the large $U$ limit.\par
We know in this non-Fermi liquid case $V_{\rm a}=V/2$ that there are singular contributions
to both $\chi_j$ and $\gamma$ as $T\to 0$. This singular temperature dependence is associated
with the derivatives of the self-energies of the vector Majorana fermions evaluated in the
limit $\omega\to 0$, and in evaluating the above expressions in this case we have to
keep a small but finite value for the temperature $T$.  We can then define a temperature dependent  factor $z_{\alpha}(T)=z(T)$ for
the vector components $\alpha=1,2,3$ via
\be z_{\alpha}(T)=\left\{1-
{{\partial\Sigma_{\alpha\alpha}(\omega)}\over{i\partial\omega}}\right\}_{\omega=0}^{-1}\ee
Then the contributions to the specific heat and susceptibility can regarded as due to
free Majorana quasiparticles each having a renormalized resonance width of 
$\tilde\Delta(T)=z(T)\Delta$. In the Wilson ratio the renormalization factor $z(T)$
cancel so the result is the same as that for the non-interacting Majorana fermions.
This is similar to what happened for the SAM ($V_{\rm a}= 0$), except that $z(T)$ in that case
was independent of $T$ and so the free quasiparticle description is the more conventional
Fermi liquid one. 
In the calculation of $\chi_j$ for the SAM  the  field $H_{12}$ acts only on the up electrons
 so that the renormalized interaction
$\tilde U$ between the quasiparticles (related to the vertex $\Gamma_{0123}(0,0,0,0)$)
 plays no role \cite{hew1}.\par
Though we have shown it is possible to interpret the results for the non-Fermi liquid
low temperature behaviour in terms of temperature dependent quasiparticles, we do not 
gain any insight into the nature of the singular terms contributing to $z(T)$ other
than from the weak coupling perturbation theory.  Also as $1/z(T)\to 0$, which follows
from the ${\rm ln}(T)$ of $\chi_j$, the quasiparticle weight vanishes at $T=0$
so in that sense the quasiparticles disappear as $T\to 0$.
In the next section we can give a more satisfactory interpretation of the results by
separating out the temperature independent contributions to $z(T)$ from the singular temperature
dependent ones. We can then define more conventional quasiparticle excitations.
The singular  scattering of these quasiparticles  is not to be intepretated as 
 a breakdown of the quasiparticle concept. It is simply due to the fact that 
 there are real quasiparticle scattering
processes which lead to real singularities in the  physical properties of this 
model as $T\to 0$.
\par

\section{Renormalized Perturbation Theory}\par

 We know that in the case of a  Fermi liquid fixed 
point there
are quasiparticle excitations from the ground state of the interacting system which are in
 one-to-one correspondence with the single particle excitations of the non-interacting system.
One of us (ACH) showed in an earlier paper \cite{hew1} that it is possible to make a 
renormalized
perturbation expansion, valid for all values of $U$, in terms of these quasiparticles about this fixed point for the O(4)
symmetric Anderson model (with a generalization to translationally invariant systems in Ref.\
\cite{hew2}), where the exact results for the charge and spin susceptibilities at $T=0$ are 
obtained from the lowest
order (tadpole) diagram and the exact impurity contribution to the conductivity as $T\to 0$
 from the second order diagrams. 
For $U=0$ in the O(3) model we have independent particle 
excitations from the ground state. We now look into the possibility of defining quasiparticles
and of deriving
a renormalized perturbation expansion for the O(3) model in the non-Fermi liquid
case $V_{\rm a}=V/2$, such that the lowest order diagrams give  exact results 
in the limit $T\to 0$. We first of all rewrite the self-energy of the Majorana
fermion retarded Green's functions in the form,
\be \Sigma_{\alpha\alpha}(\omega)=\Sigma^{(r)}_{\alpha\alpha}(\omega)+\Sigma^{(s)}_{\alpha\alpha}(\omega),\ee
where $\Sigma^{(s)}_{\alpha\alpha}(\omega)$ is the part of the self-energy at $T=0$
which gives a contribution to the derivative of $\Sigma_{\alpha\alpha}(\omega)$
which diverges as $\omega\to 0$, and $\Sigma^{(r)}_{\alpha\alpha}(\omega)$
is the remaining part. Note that as we are using retarded Green's functions here
these are in terms of real frequency variables $\omega$ rather than imaginary ones used
in the previous section, where we used the  thermal Green's functions 
(they are related by $G^{\rm ret}_{\alpha\alpha}(i\omega_n)=G_{\alpha\alpha}(\omega_n)$) This latter term we write in the form,
\be \Sigma^{(r)}_{\alpha\alpha}(\omega)=\Sigma^{(r)}_{\alpha\alpha}(0)+\omega
\Sigma'^{(r)}_{\alpha\alpha}(0)+\Sigma^{\rm (rem)}_{\alpha\alpha}(\omega).\ee
Substituting this result in the corresponding Green's function, and 
using the fact that from particle-hole symmetry $\Sigma^{(r)}_{\alpha\alpha}(0)=0$,
we find the interacting Green's function can be written in the form,
\be G^{\rm ret}_{\alpha\alpha}(\omega)={\bar z_{\alpha}\over {\omega+i\bar\Delta_{\alpha}-
\bar \Sigma^{(r)}_{\alpha\alpha}(\omega)-\bar \Sigma^{(s)}_{\alpha\alpha}(\omega)}},\ee
where
\bea
\bar\Delta_{\alpha}=\bar z_{\alpha}\Delta_{\alpha},\quad \bar \Sigma^{(r)}_{\alpha\alpha}(\omega)=
\bar z_{\alpha}\Sigma^{\rm (rem)}_{\alpha\alpha}(\omega),
\nonumber \\
\bar\Sigma^{(s)}_{\alpha\alpha}(\omega)=
\bar z_{\alpha}\Sigma^{(s)}_{\alpha\alpha}(\omega),
\eea
and $\bar z_{\alpha}=1/(1-\Sigma'^{(r)}_{\alpha\alpha}(0))$. We can rescale the Majorana field to absorb the 
$\bar z_{\alpha}$ factor, and the corresponding Green's function without this factor will
describe the corresponding quasiparticle Green's function. Following a prescription similar
to that used for the O(4) model we can rewrite the Hamiltonian in the form,
\be H=\bar H +H_{\rm c},\label{rh}\ee
where $H$ is the original or "bare" Hamiltonian, as given in Eq.\ (\ref{hamm}), $\bar H$
is a Hamiltonian of the same form where all parameters and fields have been replaced by the
corresponding quasiparticle ones (indicated by a bar), and $H_c$ consists of the remaining terms,
known as counter terms. Expressions can be derived for the counter terms, which correspond to 
hybridization with the impurity and an on-site interaction term. However, following a modified
form of the standard renormalization prescription, they can be determined by the conditions,
\bea
  A&:&\quad\quad \bar\Sigma^{(r)}_{\alpha\alpha}(0)=0,
\nonumber \\
  B&:&\quad\quad \bar\Sigma'^{(r)}_{\alpha\alpha}(0)=0,
\nonumber \\
  C&:&\quad\quad \bar U=(\bar z_0 \bar z_1 \bar z_2 \bar z_3)^{1/2}\Gamma_{0123}(0,0,0,0),
\eea
which are applied at $T=0$. These conditions express the fact the quasiparticles and the interaction term is fully renormalized
with respect to the non-singular contributions. One can now develop a renormalized perturbation
expansion for the self-energies $\bar\Sigma^{(r)}_{\alpha\alpha}(\omega)$ and 
$\bar\Sigma^{(s)}_{\alpha\alpha}(\omega)$ in which one expands both in the interaction term
in $\bar H$ and all the counter terms in $H_{\rm c}$. These are organized order by order
in powers of $\bar U$. The condition B expresses the requirement
that the quasiparticle Green's function has weight 1, so the renormalized fields have to be
scaled appropriately  by a change $\delta \bar z_{\alpha}$, which is also a counter term but it does
not occur explicitly in (\ref{rh}). If, however, we use the functional integral formalism
 then the equation corresponding to (\ref{rh}) is
\be  S=\bar S +S_{\rm c},\label{sh}\ee
where $S$ is the action, $\bar S$ the renormalized action and $S_{\rm c}$ the counter terms.
The counter term contribution in this case  contains an explicit term in $\delta \bar z_{\alpha}$, 
arising from the term in (\ref{fia}) involving the derivative of the fields with respect to 
$\tau$. In this respect the functional integral  formulation of the renormalized perturbation
expansion is the more natural one and corresponds to the standard
renormalization procedure for the $\phi^4$ field theory \cite{ft}.\par 
The renormalized perturbation prescription corresponds simply to a reorganization of the original
perturbation theory, no terms have been neglected. Up to this point we have been
quite general and, if we have no singular contributions, the renormalized expansion
is the same as that for the O(4) model with $V_{\rm a}=0$, which describes an expansion
about a Fermi liquid fixed point. The prescription here, however, generalizes the expansion
about the Fermi liquid fixed point to the case of any $V_{\rm a}$ such that
 $V_{\rm a}\ne V/2$.
We now consider the non-Fermi liquid case, $V_{\rm a}= V/2$. In this case we
simplify the notation and
 drop the
subscripts for $\alpha=1,2,3$ and retain them only for $\alpha=0$. 
The results for the specific heat coefficient $\gamma$ and the susceptibility $\chi_j$
at $T=0$ are due to the non-interacting quasiparticle system described by $\bar H(\bar U=0)$,
\be \gamma={{\pi k_{\rm B}^2}\over {2\bar\Delta}},
\quad\quad \chi_j= {{(g_j\mu_{\rm B})^2}\over {\pi\bar\Delta}}. \label{qpr}\ee
These give a Wilson $\chi/\gamma$ ratio of ${8\over 3}$.
There are no regular corrections to these results at $T=0$ (in zero field $H_{12}=0$) 
arising from the renormalized
expansion in powers of $\bar U$.  This is a consequence of the Ward identity we derived
in the previous section. We can write the wavefunction renormalization factor $z(T)$
used there as $z(T)=\bar z +z^{(s)}(T)$ 
where $\bar z=1/(1-\Sigma^{\prime (r)}(0))$ and
$z^{(s)}(T)$  is the remaining part which contains $\ln T$ terms. 
The result from the non-interacting
quasiparticles as defined above takes account of all but the singular terms.
The singular contributions
to the susceptibility 
as we saw in the previous section, arise from the diagrams which can be severed into separate diagrams by cutting a pair of lines,
one of which must correspond to the $\alpha=0$ propagator.
The results  for  the second order renormalized diagrams for these singular contributions are
\bea
 \gamma^{s}&=&-{{\pi k_{\rm B}^2}\over {2\bar\Delta}}\left ({\bar U\over {\pi\bar\Delta}}\right)^2
{\rm ln}\left({{k_{\rm B}T}\over{\bar\Delta}}\right),
 \\
\chi^{s}_j&=&-{{(g_j\mu_{\rm B})^2}\over {\pi\bar\Delta}}\left ({\bar U\over {\pi\bar\Delta}}\right)^2
{\rm ln}\left({{k_{\rm B}T}\over{\bar\Delta}}\right).\label{rr}
\eea
These are the only singular terms as $T\to 0$, as higher order weak coupling diagrams 
to order ${\rm ln}(T)$, only contribute to the non-singular renormalized four vertex 
and so are already included in $\bar U$. There are no counter term diagrams to
take into account for the singular terms.
The higher order terms in ${\rm ln}(T)$ cancel
from the same arguments given in the discussion of the weak coupling theory.
In the limit $U\to 0$, $\bar U\to U$, $\bar \Delta\to\Delta$, and we recover the
weak coupling result. However, the renormalized theory is valid for all
values of $U$ and so if we take $U\gg \pi\Delta$, in which parameter range we
can map the model into the $\sigma$-$\tau$ model then the leading
order corrections to the specific heat and susceptibility as $T\to 0$
are given by the same diagrams but now there is only one low energy scale,
the Kondo temperature $T_{\rm K}$, so $\bar U\sim \pi\bar\Delta$. If we put together the regular and singular terms we again get a Wilson 
$\chi/\gamma$ ratio of ${8\over 3}$ independent of $U$ (and $\bar U$), as we would expect from
the Ward identity argument given in the previous section. If we compare these results with
those derived via the Bethe ansatz for the two channel Kondo model \cite{aj} then the results are
the same if $\bar U/\pi\bar\Delta=1$ with $\bar\Delta=T_{\rm K}$ (which also holds for the symmetric 
Anderson model in the Kondo limit \cite{hew1}).\par
It is only
in the large $U$ limit  ($U\gg \pi\Delta$), when the isopin fluctuations of the
impurity are suppressed, that  we can identify $\chi_j$ with the spin susceptibility
of the impurity. For  $U>\pi\Delta$,  we cannot calculate the spin susceptibility
of the impurity from just the low order terms in the renormalized perturbation expansion.
This is because the spin is not conserved and an expression for the spin susceptibility
cannot be derived simply from a knowledge of the low lying energy levels. The same
holds true in the NRG approach. The calculation of $\chi_j$ via the NRG is relatively
straightforward because the total spin plus isospin is conserved and the susceptibility
can be derived in terms of the energy eigenvalues. The  spin
susceptibility $\chi_s$ is much more difficult to calculate because it requires 
a knowledge of  matrix elements which  have to be calculated and updated at each
step in the NRG calculation.  There is another way of looking at this difficulty.
We can calculate the spin susceptibility if we know the low lying levels of the 
system in the presence of a magnetic field which couples only  to the impurity
spin. In principle  these levels could be derived from an effective
Hamiltonian which describes the system near the low energy fixed point  in the presence
of a magnetic field. However,
there would be another renormalized parameter in this Hamiltonian associated
with the coupling of the spin to the field, which we can interpret as
a renormalized g-factor. We could calculate the spin susceptibility but
the g-factor would be unknown. Similar arguments apply to the calculation of
the charge susceptibility $\chi_c$ of the impurity. 
\par
There is no problem in calculating $\chi_s$ and $\chi_c$ for the O(4)
model for $V_{\rm a}=0$ as both  charge and spin
are conserved. Exact results are obtained for these quantities (at $T=0$)
from the renormalized perturbation theory to first order in $\bar U$ \cite{hew1}.
The higher order contributions can be shown to cancel as a result
of a Ward identity.  Coleman and Schofield \cite{cs} have used
the renormalized perturbation theory to first order in $\bar U$ to
calculate 
$\chi_s$ and $\chi_c$ for the model with $V_{\rm a}\ne  0$ but 
$V_{\rm a}\ne  V/2$  and deduced a value for the Wilson ratio.
 However, as the spin and charge are no longer conserved performing
the calculation only to first order in $\bar U$ in this case cannot be exact.
Their result for the Wilson ratio $\chi_s/\gamma$ 
was found to be in satisfactory agreement with that given by the Bethe ansatz
for the anisotropic two channel model and to provide an interpolation over the 
whole parameter range (though it differs by a factor of two from the Bethe ansatz result
in the limit $(J_1-J_2)\ll J_1$). \par
We can give an alternative argument to that of Coleman and Schofield using 
renormalized perturbation theory
to calculate both $\chi$ and $\gamma$ in the  limit $\Delta_0\ll\Delta$, $U\gg \pi\Delta$. 
 In this limit logarithmic terms
in $\Delta_0\ll\Delta$ appear associated with the same diagrams that give the logarithmic
in $T$ contributions for $\Delta_0=0$. We can  reorganise the renormalized perturbation
theory for  $\Delta_0\ll\Delta$ 
so that $\Sigma^{(s)}_{\alpha\alpha}(\omega)$ refers to the diagrams that give 
logarithmic terms in  $\Delta_0$ as
$\Delta_0\to 0$.  We  calculate
asymptotically the leading order contributions to the specific heat and susceptibility 
in a similar way to the above calculation, 
\be \gamma={{\pi k_{\rm B}^2}\over 6}
\left({3\over {\bar\Delta}}\left\{1-\left({\bar U\over {\pi\bar\Delta}}\right)^2{{{\rm ln}(\bar\Delta_0/\bar\Delta)}
\over{\pi\bar\Delta}}\right\}
+{1\over {\bar\Delta_0}}
\right),\label{ga}\ee
\be \chi_j=(g_j\mu_{\rm B})^2\left({1\over {\pi\bar\Delta}}- 
\left ({\bar U\over {\pi\bar\Delta}}\right)^2{{{\rm ln}(\bar\Delta_0/\bar\Delta)}
\over{\pi\bar\Delta}}\right).\label{ca}\ee
In the specific heat term the most singular contribution as $\bar\Delta_0\to 0$
is the one proportional to $\bar\Delta_0^{-1}$ and so dominates in the Wilson ratio in this
limit. We found earlier that the results in the Kondo limit $U\gg\pi\Delta$
($\chi_j\to\chi_s$) correspond to the two channel Kondo model
for $\bar U/\pi\bar\Delta=1$ so we can use this value in the coefficient of the 
leading singular term. 
 The Wilson ratio then gives $-8(\bar\Delta_0/\bar\Delta)
{\rm ln}(\bar\Delta_0/\bar\Delta)$ as $(\bar\Delta_0/\bar\Delta)\to 0$ in complete agreement with
the Bethe ansatz result for the anisotropic two channel model in this limit \cite{aj}
where we identify  $(\bar\Delta_0/\bar\Delta)$ as the ratio of the renormalized energy scales
$T_{\rm a}/T_{\rm i}$. The Wilson ratio based on  (\ref{ga}) and (\ref{ca}) will
also be correct for the O(4) model which corresponds to the limit $\bar\Delta_0\to \bar\Delta$.
 The logarithmic terms then vanish and we recover
a value of 2  obtained earlier ($\bar U/\pi\bar \Delta=1$ in this limit also but
this factor plays no role in the result as it multiplies the log terms only). Hence this result
is in complete agreement  with those derived from the Bethe ansatz in both limits,
and also provides an interpolation between them.\par

\section{Conclusions}
The two models we have been considering here were both introduced as a way of gaining insight
into 
the nature of the low energy fixed point for the isotropic and anisotropic two
channel Kondo models. Our results  confirm  that there are very strong similarities in the results
for these compactified models (in the localized or Kondo limit) and the two channel Kondo
model, and that the low temperature behaviour of the compactified models can be given a
simple interpretation in terms an effective Majorana fermion model at the low energy fixed point.
A consistent picture of the isotropic and anisotropic fixed points  emerges 
from the perturbational, numerical renormalization
group, conformal field theory, and renormalized perturbation theory approaches. There are, however,
some important differences which have not been brought out in previous work on these models.\par
In the NRG results for the isotropic two channel model \cite{cln,pc} it has been found that as the fixed 
point is approached that  there is no change of the spectrum as the number sites $N$ included 
in each of the two chains changes from $N$ to $N+1$. This is in contrast to the equivalent results for the single
channel Kondo model where the spectrum of low lying many-body levels varies as to whether
$N$ is odd or even, and where there are two equivalent fixed points corresponding to
a transformation $N\to N+2$, one for $N$ odd and the other for $N$ even. Superficially
it appears that there is also no change in the spectrum in the NRG results for the compactified
models for the isotropic case as $N\to N+1$ as the fixed point is approached. However this
is not the case. The degeneracies of the excitations depend on whether $N$ is odd or even.
The levels and the degeneracies can both be explained on the basis of free vector and scalar
Majorana fermion chains with different boundary conditions  as described in Sec.\ IV. 
When $N$ is even the many-body excitations correspond to those given in Table VII, which
are built up from those of a scalar Majorana fermion chain with periodic boundary conditions
and a vector one with antiperiodic boundary conditions. When $N$ is odd they correspond to 
those given in Table
VIII where the boundary conditions on the scalar and vector components have been interchanged.
As in the one channel Kondo case, these are alternative situations and cannot be superimposed
on each other. However, to reproduce the excitation spectrum as found in the NRG calculations
for the two channel model these alternative situations
(refered to as sector I and sector II) have to be combined. The degeneracies
of the many-body excitations of the two channel model can be explained by including the extra uncoupled 
degrees of freedom in the form of an extra non-interacting fermion chain.
 As this is uncoupled from the impurity like the scalar Majorana fermion chain
it is taken to have the same boundary conditions as the scalar part. With this model
both the many-body energies and the degeneracies of the NRG spectrum at the fixed point
can be reproduced. Only the ground state degeneracy, which only acts as a multiplicative factor
for the degeneracies of the excited states, has to be reduced as compared to the
compactified model.

The single particle levels of the combined chains with the appropriate
boundary conditions are shown in Fig.\ 10 (for comparison with the spectra
of the compactified model see Fig.\ 11).
 The resulting many-body energies with the degeneracies
are given in Table IX for sector I and Table X for sector II. 
They agree with those given in the NRG calculations of Cragg {\it et al.} \cite{cln}
and Pang and Cox \cite{pc} and the conformal field theory results of Affleck {\it et al.}\cite{al,alpc}.
This spectrum has also been obtained for the two channel Kondo model from
single particle excitations corresponding to different boundary conditions by Ye \cite{ye1} using
the bosonization approach. Ye in a recent paper  has also constructed the excitation spectrum for the fixed
point of the compactified models in a similar way \cite{ye2}. His results differ from ours
as he has the two sectors, which we identify with $N$ odd and $N$ even, combined as in the
two channel case. This is not consistent with the NRG results.\par
Though  Affleck and Ludwig, in their conformal field theory for the two channel Kondo model,
 could explain  the excitation spectrum found in the NRG calculations at the fixed point, 
they could not
derive the many-body states from a single particle picture with modified boundary conditions.
Nor could they obtain an explicit form for the leading irrelevant operator at the fixed point
in terms of local operators. We have shown, however, that both these can be
achieved within the conformal field theory approach for the $\sigma$-$\tau$ model.
 In Sec.\ III we found the same two
separate sectors of excitations, built up from free  Majorana chains with different
boundary conditions as in the NRG approach, and the same local operator as in the NRG 
was found for the
leading irrelevant interactions
using very similar arguments to those used for the single channel Kondo model. With
these insights from the CFT of the $\sigma$-$\tau$ model
it might prove possible to get a deeper understanding of the two channel
conformal field theory, and to understand why the sectors corresponding to Tables VII and VIII
have to be combined for that system.\par
At this point it would seem appropriate to compare and contrast our results
with those of previous papers on these models. We agree with the work of
Coleman {\it et al.} \cite{cit} that the fixed point of the isotropic $\sigma$-$\tau$
model does correspond to strong coupling and we agree with the form of the leading
irrelevant interaction. However we find that $D/J$ is not a good expansion
parameter as $(D/J)^2$ corresponds to
 $\bar U/\pi\bar\Delta$ which is  of order unity.
 We have also been able to resolve the problems of the higher
order logarithmic terms that were  found in our earlier work on this model (G-M.Z., A.C.H.) \cite{zh} in the
weak coupling expansion for the O(3) symmetric Anderson model ($V_0=0$). Explicit
cancellation of these higher order log terms to the vertex $\Gamma_{0123}(0,0,0,0)$ and
the susceptibility 
$\chi_j$ has been demonstrated up to fourth order, and we believe this can be generalized to
all orders in $U$. This means that the marginal Fermi liquid fixed point is marginally stable rather
than marginally unstable.\par
As the nature of the fixed point found in the NRG calculations
is always independent of $U$, we have found it  natural to describe the
fixed point as a renormalized form of the Anderson model. This is similar to the case
of the standard Anderson impurity model where the fixed point always corresponds to
a Fermi liquid fixed point, whatever the value of $U$, and the low temperature behaviour
can be described by a renormalized version of the same model. In the strong correlation regime
the parameters of the renormalized model are not independent. They  depend on the
single energy scale set by the Kondo temperature $T_{\rm K}$. We find similar results
for the marginal Fermi liquid fixed point of the O(3) model. The main difference
with the standard Anderson model is that there is a singular scattering mechanism
of the quasiparticles which leads to real singularities in some of the properties
of the model, such as the susceptibility, as $T\to 0$. This can be described by a 
 modified form of the
 renormalized perturbation theory that was developed for the O(4) model \cite{hew1},
in which the low temperature properties are given exactly from the diagrammatic
expansion up to second order.\par
For the O(3) Anderson model with $0<V_0<V$ in the strong correlation regime the
renormalized model has two energy scales rather than one, as does the corresponding
$\sigma$-$\tau$ model and the anisotropic two channel model \cite{aj}. Apart from the fact
that there are two independent energy scales the fixed point behaviour is similar
to that for the O(4) model and could be described as a Fermi liquid fixed point
(as in \cite{fgn}).
As the spin and charge are not conserved for this model we found that the first order
renormalized perturbation theory for the spin and charge susceptibilities,
as used by Coleman and Schofield \cite{cs}, is not exact
as it is in the case of the O(4) model (where they are conserved). However,
we found that it is possible within this approach to calculate the Wilson
ratio asymptotically in the limit $V_0\to 0$ which gives a result in complete
agreement with that from the Bethe ansatz for the anisotropic 
Kondo model in the same limit.\par
The O(3) symmetric Anderson model is of  interest apart from its relation to the 
two channel Kondo model as it displays a marginal Fermi liquid fixed point for 
$V_0=0$. The resistivity for this model does not correspond to the
resistivity or the two channel Kondo model and consequently has a different
temperature dependence. In the lowest order perturbation theory it
was shown earlier \cite{zh} that to lowest order in $U$ it is linear in $T$ term,
as in the marginal Fermi liquid theory. The dynamics of this model over the
full parameter range is presently being calculated using the NRG \cite{bbh}.\par
\bigskip
\noindent{\bf Acknowledgment}\par
\bigskip
We thank  Y. Chen, J. von Delft and Th. Pruschke  for helpful conversations.
We are grateful to the EPSRC for the support of a research grant
(grant No.\ GR/J85349), and to the DFG (grant No.\ Bu965-1/1)
for a research fellowship for one of us (RB).\par

\begin{table}
\begin{center}
\caption{Many-body excitation energies and their
  corresponding degeneracies for a single 
  Majorana fermion chain with periodic boundary conditions.
  The single particle spectrum is given in Fig.\ 1(a).}
\begin{tabular}[t]{cccc}
   $E_{\rm ex}/(\pi v_{\rm F}/l)$  & $\sum_k n_k \epsilon_k $
       & $dg$   & total $dg$\\
  \hline
   0 & 0 & 2 & 2 \\
  \hline
   1 &$\epsilon_{1}$ & 2 & 2 \\
  \hline
   2 &$\epsilon_{2}$ & 2 & 2 \\
  \hline
   3 &$\epsilon_{3}$ & 2 &  \\
     &$\epsilon_{1}+\epsilon_{2}$ & 2 & 4 \\
  \hline
   4 &$\epsilon_{4}$ & 2 &  \\
     &$\epsilon_{1}+\epsilon_{3}$ & 2 & 4 \\
\end{tabular}
\end{center}
\end{table}

\begin{table}
\begin{center}
\caption{Many-body excitation energies and their
  corresponding degeneracies for a single 
  Majorana fermion chain with antiperiodic boundary conditions.
  The single particle spectrum is given in Fig.\ 1(b).}
\begin{tabular}[htb]{cccc}
   $E_{\rm ex}/(\pi v_{\rm F}/l)$  & $\sum_k n_k \epsilon_k $
       & $dg$   & total $dg$\\
  \hline
   0 & 0 & 1 & 1 \\
  \hline
   1/2 &$\epsilon_{1/2}$ & 1 & 1 \\
  \hline
   3/2 &$\epsilon_{3/2}$ & 1 & 1 \\
  \hline
   2 &$\epsilon_{1/2}+\epsilon_{3/2}$ & 1 & 1 \\
  \hline
   5/2 &$\epsilon_{5/2}$ & 1 & 1 \\
  \hline
   3 &$\epsilon_{1/2}+\epsilon_{5/2}$ & 1 & 1 \\
  \hline
   7/2 &$\epsilon_{7/2}$ & 1 & 1 \\
  \hline
   4 &$\epsilon_{1/2}+\epsilon_{7/2}$ & 1 &  \\
     &$\epsilon_{3/2}+\epsilon_{5/2}$ & 1 & 2 \\
\end{tabular}
\end{center}
\end{table}

\begin{table}
\begin{center}
\caption{Many-body excitation energies and their
 corresponding degeneracies for a vector
  Majorana fermion chain with periodic boundary conditions.
 The single particle spectrum is given in Fig.\ 2(a).}
\begin{tabular}[t]{cccc}
   $E_{\rm ex}/(\pi v_{\rm F}/l)$  & $\sum_k n_k \epsilon_k $
       & $dg$   & total $dg$\\
  \hline
   0 & 0 & 4 & 4 \\
  \hline
   1 &$\epsilon_{1}$ & 12 & 12 \\
  \hline
   2 &$\epsilon_{2}$ & 12 &  \\
    &$2\epsilon_{1}$ & 12 & 24 \\
  \hline
   3 &$\epsilon_{3}$ & 12 &  \\
     &$3\epsilon_{1}$ & 4 & 52 \\
     &$\epsilon_{1}+\epsilon_{2}$ & 36 &  \\
  \hline
   4 &$\epsilon_{4}$ & 12 &  \\
     &$\epsilon_{1}+\epsilon_{3}$ & 36 & 84 \\
     &$2\epsilon_{1}+\epsilon_{2}$ & 36 &  \\
\end{tabular}
\end{center}
\end{table}

\begin{table}
\begin{center}
\caption{Many-body excitation energies and their
  corresponding degeneracies for a vector 
  Majorana fermion chain with antiperiodic boundary conditions.
  The single particle spectrum is given in Fig.\ 2(b).}
\begin{tabular}[t]{cccc}
   $E_{\rm ex}/(\pi v_{\rm F}/l)$  & $\sum_k n_k \epsilon_k $
       & $dg$   & total $dg$\\
  \hline
   0 & 0 & 1 & 1 \\
  \hline
   1/2 &$\epsilon_{1/2}$ & 3 & 3 \\
  \hline
   1  &$2\epsilon_{1/2}$ & 3 & 3 \\
  \hline
   3/2 &$\epsilon_{3/2}$ & 3 &  \\
       &$3\epsilon_{1/2}$ & 1 & 4 \\
  \hline
   2 &$\epsilon_{1/2}+\epsilon_{3/2}$ & 9 & 9 \\
  \hline
   5/2 &$\epsilon_{5/2}$ & 3 &  \\
    &$2\epsilon_{1/2}+\epsilon_{3/2}$ & 9 & 12 \\
  \hline
   3 &$\epsilon_{1/2}+\epsilon_{5/2}$ & 9 &  \\
     &$3\epsilon_{1/2}+\epsilon_{3/2}$ & 3 & 12 \\
  \hline
   7/2 &$\epsilon_{7/2}$ & 3 &  \\
     &$2\epsilon_{1/2}+\epsilon_{5/2}$ & 9 & 21 \\
     &$\epsilon_{1/2}+2\epsilon_{3/2}$ & 9 &  \\
\end{tabular}
\end{center}
\end{table}

\begin{table}
\caption{Finite-size spectrum of the non-interacting case for PBC (with
degenerate ground state). Compared with the APBC case, the ground state energy
shifts by $1/4$ unit. $dg$ is the degeneracy of the energy level. }
\begin{center}
\begin{tabular}{cccccc}
$E_{\rm ex}/(\pi v_{\rm F}/l)$ & $j$ & $\Delta_{\rm Ising}$ &
 $n_{\rm vec}+n_{\rm sc}$ & $dg$ & total $dg$ \\ \hline
     0   &     1/2    &   1/16   &    0    &     4   &   4   \\ 
\hline
     1   &     1/2    &   1/16   &    1    &    16   &   16  \\
\end{tabular}
\end{center}
\end{table}

\begin{table}
\caption{Finite-size spectrum of the non-interacting case for APBC (with
non-degenerate ground state).}
\begin{center}
\begin{tabular}{cccccc}
$E_{\rm ex}/(\pi v_{\rm F}/l)$ & $j$ & $\Delta_{\rm Ising}$ & 
 $n_{\rm vec}+n_{\rm sc}$ & $dg$ & total $dg$ \\ \hline
    0    &    0     &    0     &     0    &   1    &     1  \\ 
\hline
      &    0     &    1/2   &     0    &   1    &           \\
  1/2 &    1     &     0    &     0    &   3    &    4      \\
\hline
      &    1     &    1/2   &     0    &   6    &           \\
  1   &    0     &     0    &     1    &   4    &    10     \\
\end{tabular}
\end{center}
\end{table}

 \begin{table}
\caption{Finite-size spectrum for $J=J^*$
  corresponding to the vector part with APBC and scalar part with PBC. }
\begin{center}
\begin{tabular}{cccccc}
$E_{\rm ex}/(\pi v_{\rm F}/l)$ & $j$ & $\Delta_{\rm Ising}$ &
 $n_{\rm vec}+n_{\rm sc}$ & $dg$ & total $dg$ \\ \hline
  0   &    0     &     1/16    &   0    &    2    &   2   \\
\hline
  1/2 &    1     &     1/16    &   0    &    6    &   6   \\
\hline
   1  &    0     &     1/16    &   1    &    8    &   8   \\
\end{tabular}
\end{center}
\end{table}

\begin{table}
\caption{Finite-size spectrum for $J=J^*$
  corresponding to the vector part with PBC and scalar part with APBC. }
\begin{center}
\begin{tabular}{cccccc}
$E_{\rm ex}/(\pi v_{\rm F}/l)$ & $j$ & $\Delta_{\rm Ising}$ &
 $n_{\rm vec}+n_{\rm sc}$ & $dg$ & total $dg$ \\ \hline
  1/8   &    1/2    &    0     &     0      &    2     &   2 \\
\hline
  5/8   &    1/2    &   1/2    &     0      &    2     &   2 \\
\hline
  9/8   &    1/2    &   0      &     1      &    8     &   8  \\
\end{tabular}
\end{center}
\end{table}

\begin{table}
\caption{Many-body excitation energies and their
  corresponding degeneracies for sector I of the  two channel Kondo model
corresponding to the single particle spectrum of Fig.\ 10(a).}
\begin{tabular}[t]{cccc}
   $E_{\rm ex}/(\pi v_{\rm F}/l)$  & $\sum_k n_k \epsilon_k $
       & $dg$   & total $dg$\\
  \hline
   0 & 0 & 2 & 2 \\
  \hline
   1/2 &$\varepsilon_{1/2}$ & 10 & 10 \\
  \hline
   1 &$\varepsilon_{1}$ & 6 &  \\
     &$2\varepsilon_{1/2}$ & 20 & 26 \\
  \hline
   3/2 &$\varepsilon_{3/2}$ & 10 &  \\
       &$\varepsilon_{1} + \varepsilon_{1/2}$ & 30 & 60 \\
       &$3\varepsilon_{1/2}$ & 20 &  \\
\end{tabular}
\end{table}

\begin{table}
\caption{Many-body excitation energies and their
  corresponding degeneracies for sector II of the  two channel Kondo model
corresponding to the single particle spectrum of Fig.\ 10(b).
The energies are measured relative to the ground state of sector I.}
\begin{tabular}{cccc}
   $E_{\rm ex}/(\pi v_{\rm F}/l)$  &  $\sum_k n_k \epsilon_k $
       & $dg$   & total $dg$\\
  \hline
   1/8 & 0 & 4 & 4 \\
  \hline
   5/8  &$\varepsilon_{1/2}$ & 12 & 12 \\
  \hline
   9/8 &$\varepsilon_{1}$ & 20 &  \\
     &$2\varepsilon_{1/2}$ & 12 & 32 \\
  \hline
   13/8 &$\varepsilon_{3/2}$ & 12 &  \\
       &$\varepsilon_{1} + \varepsilon_{1/2}$ & 60 & 76 \\
       &$3\varepsilon_{1/2}$ & 4 &  \\
\end{tabular}
\end{table}

\begin{figure}[htb]
\epsfxsize=3.2in
\epsffile{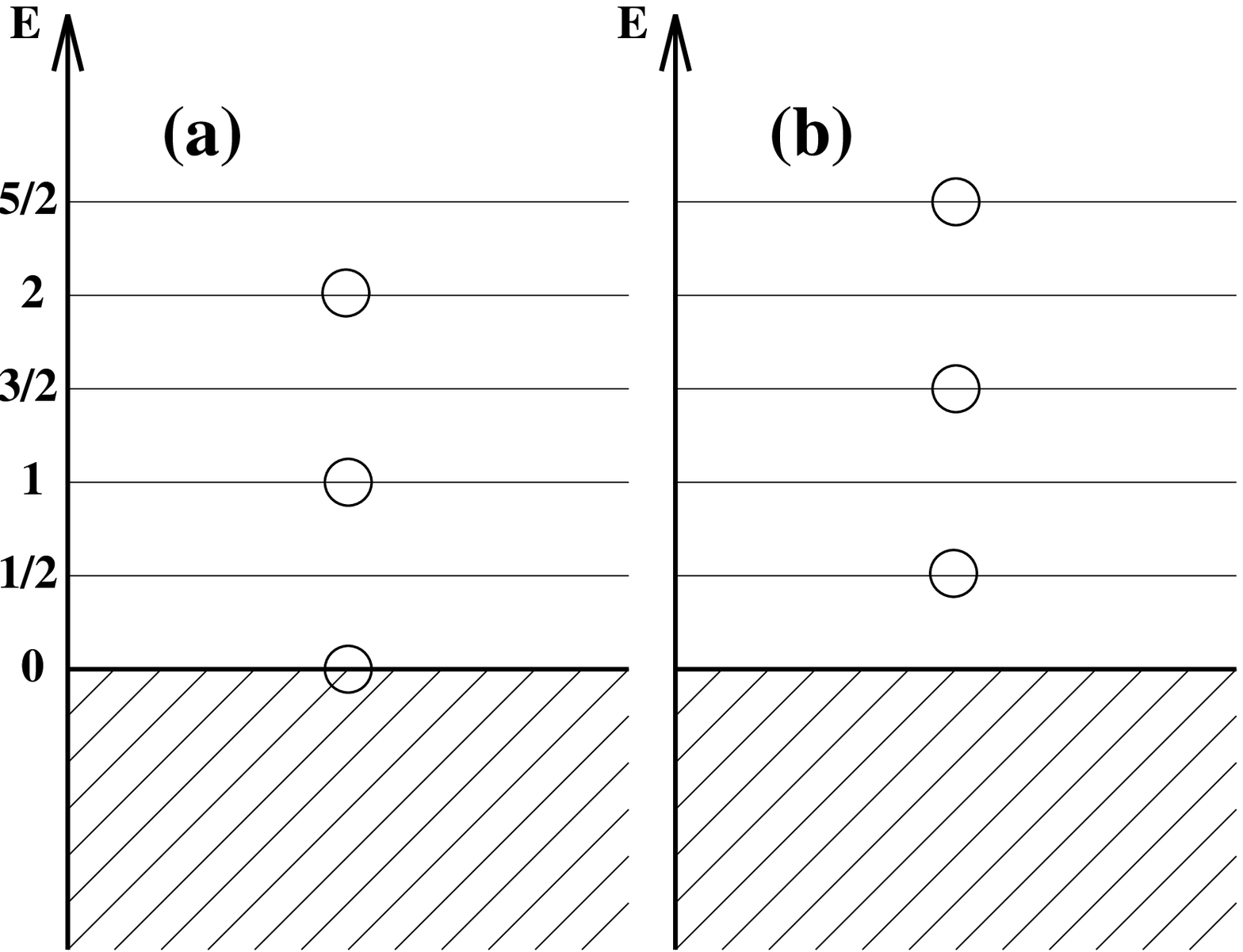}

\vspace*{0.5cm}

\caption{
Single particle spectrum of a single component 
Majorana fermion chain; (a) for periodic boundary conditions,
(b) for antiperiodic boundary conditions. Energies are give in units
of $\pi v_{\rm F}/l$.}
\end{figure}

\begin{figure}[htb]
\epsfxsize=3.2in
\epsffile{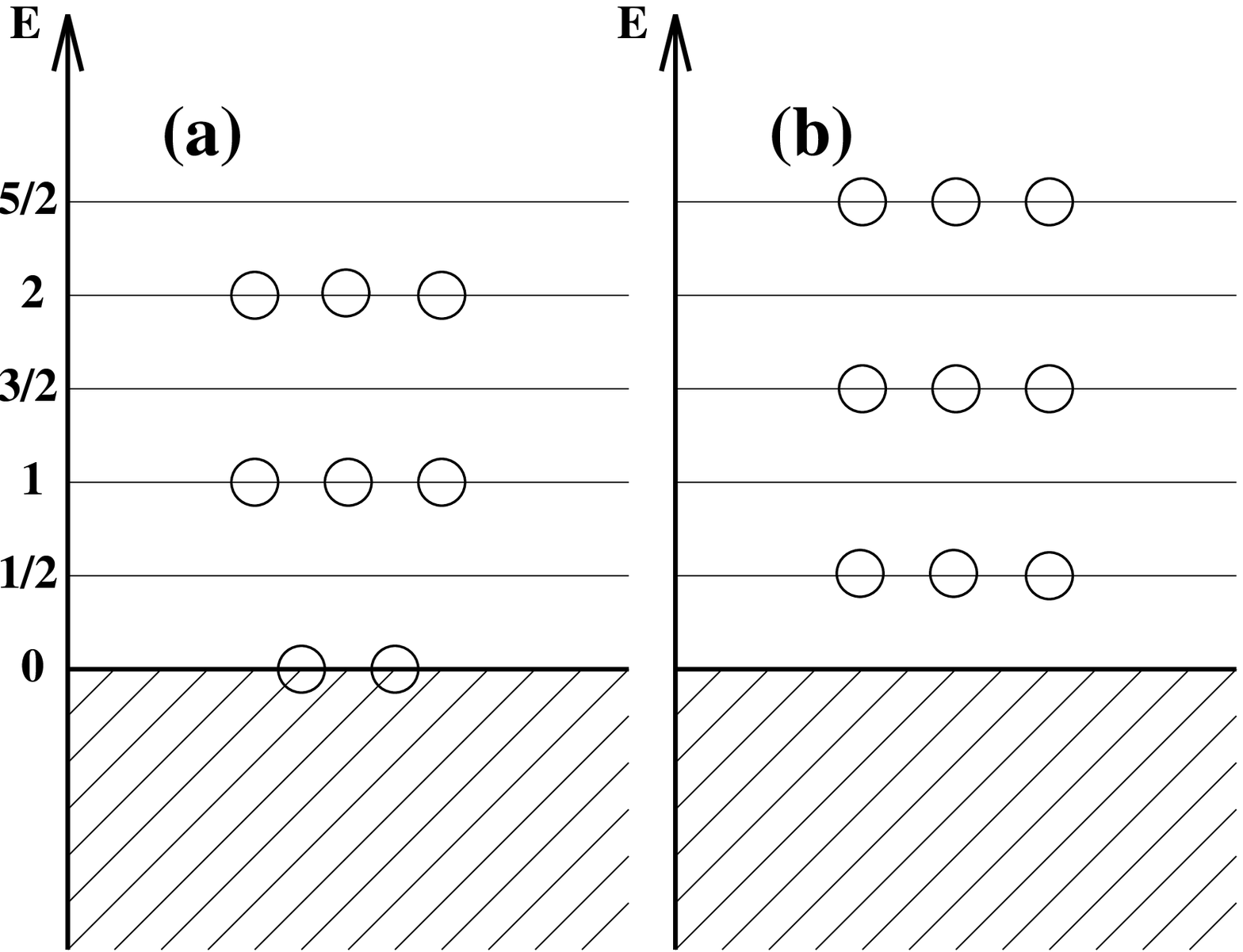}

\vspace*{0.5cm}

\caption{Single particle spectrum of a vector 
Majorana fermion chain; (a) for periodic boundary conditions,
(b) for antiperiodic boundary conditions. Energies are give in units
of $\pi v_{\rm F}/l$.}
\end{figure}

\begin{figure}[htb]
\epsfxsize=3.2in
\epsffile{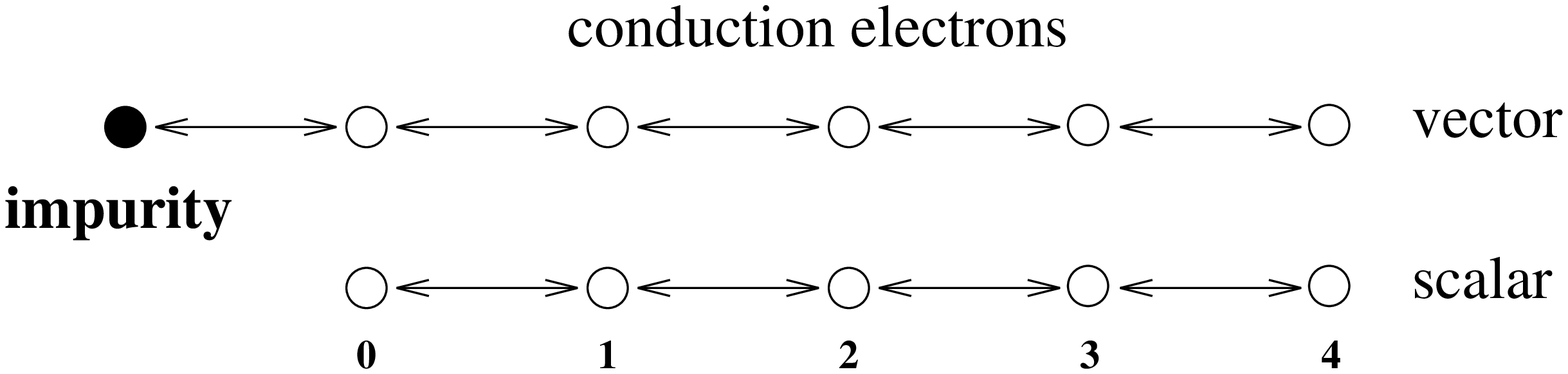}

\vspace*{0.5cm}

\caption{The separation of the model in the lattice form.   
  The upper chain represents the vector part of the conduction electrons 
  consisting of $\psi_1$, $\psi_2$, and $\psi_3$, while the lower chain 
  represents the scalar part $\psi_0$, which decouples from the impurity spin.
}

\end{figure}
\begin{figure}[htb]
\epsfxsize=3.2in
\epsffile{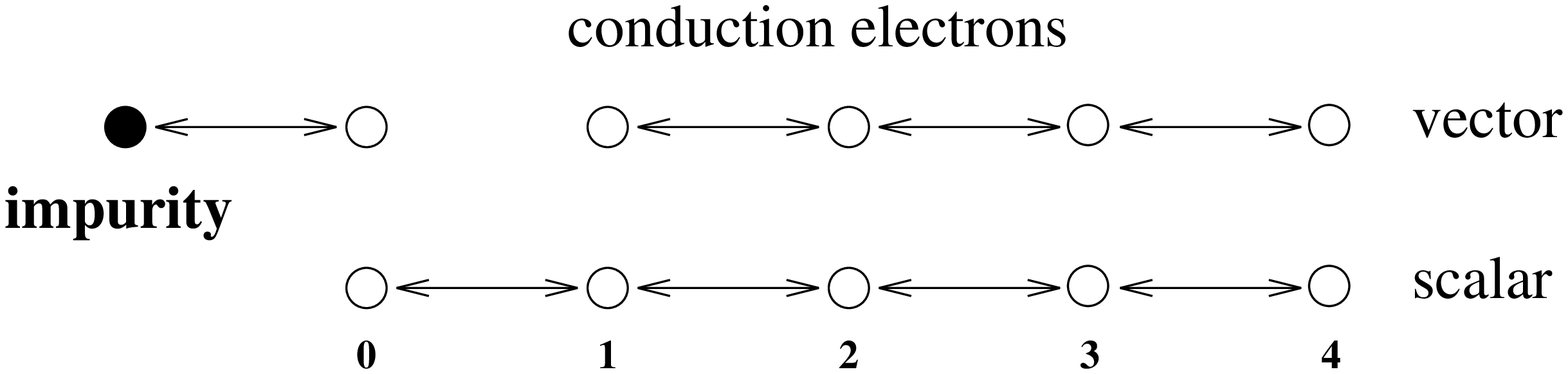}

\vspace*{0.5cm}

\caption{The picture of the strong-coupling fixed point Hamiltonian in the 
 lattice version. The impurity spin combines one electron at the nearest 
site to form a singlet bound state, and both the impurity spin and the nearest
site of the vector part  thus drop out of the problem. 
}
\end{figure}

\begin{figure}[htb]
\label{fig:flow_diagram}
\epsfxsize=3.4in
\epsffile{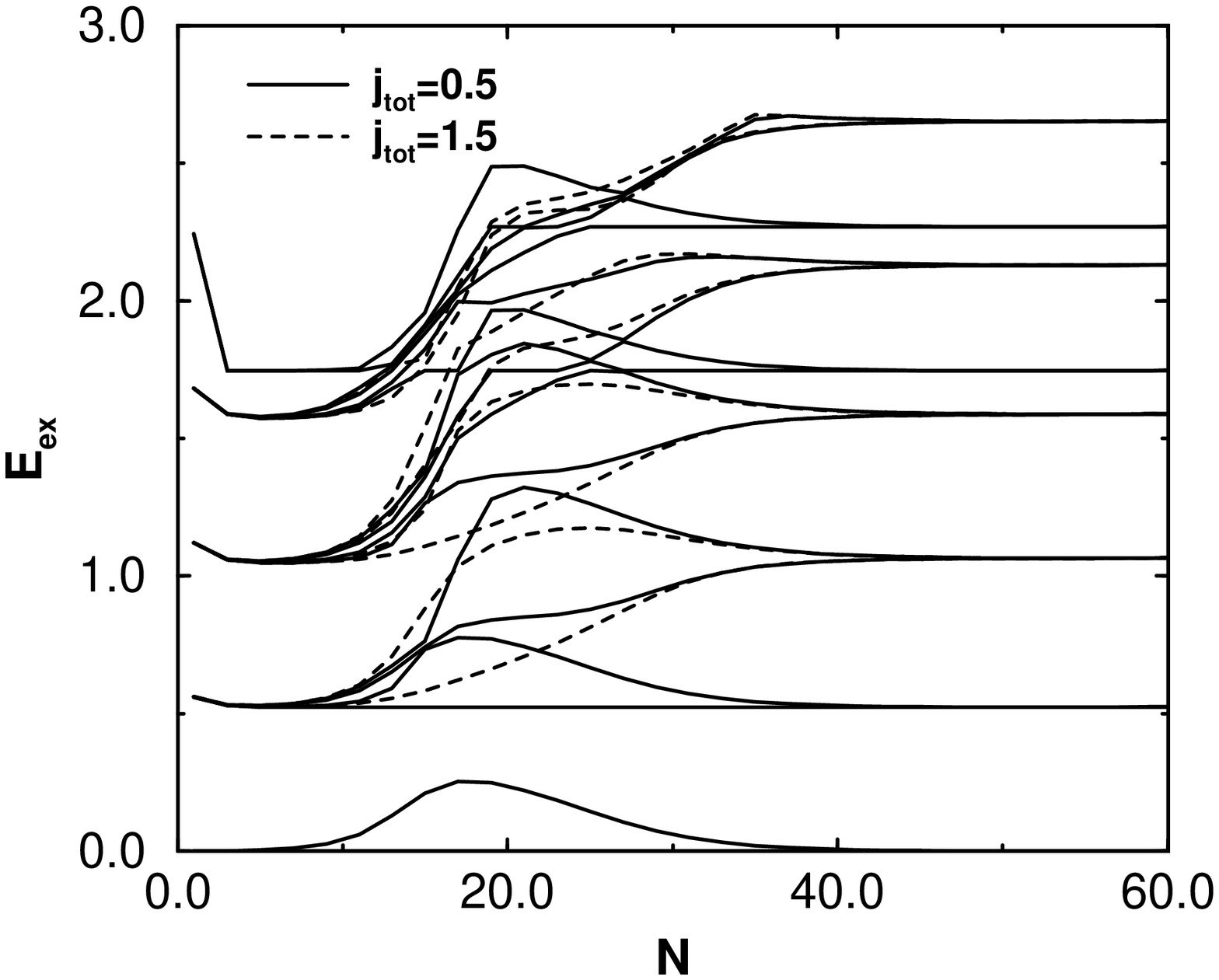}
\caption{Flow diagram for $U=0.001428$,
$V=0.01414$ and $V_{\rm a} = V/2$. Solid and dashed lines belong to
 $j_{\rm tot}=0.5$ and $j_{\rm tot}=1.5$, respectively.  The system flows
  to the non-Fermi liquid fixed point.}
\end{figure}

\newpage

\begin{figure}[htb]
\unitlength1in
\epsfxsize=3.4in
\epsffile{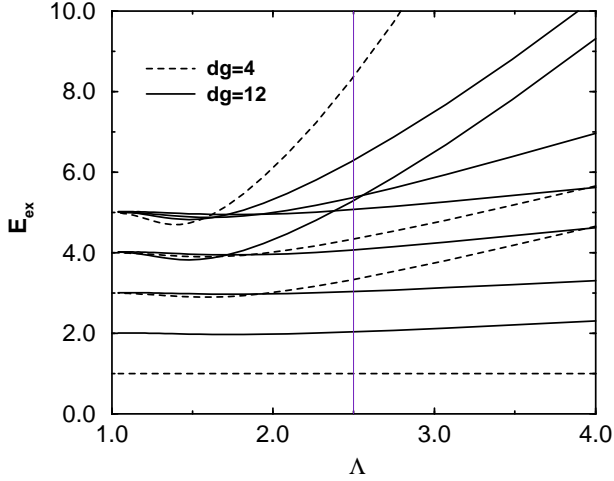}
\caption{$\Lambda$-dependence of the many-body energies $E_{\rm ex}$. The ground state
               ($E_{\rm ex}\!=\!0$) has four-fold degeneracy. Degeneracies of the excited states
                plotted here are $dg\!=\!4$ (dashed line) and $dg\!=\!12$ (solid line).
               The points where the $E_{\rm ex}(\Lambda)$ cross the vertical line at $\Lambda\!=\!2.5$
               correspond to the energies obtained from the NRG calculation.}
\end{figure}

\begin{figure}[htb]
\epsfxsize=3.2in
\epsffile{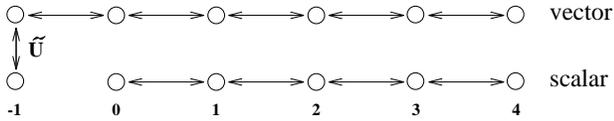}

\vspace*{0.5cm}

\caption{Structure of the non-Fermi liquid fixed point. $\tilde{U}$ couples the
vector part at site -1 to an additional Majorana-fermion which itself is not
coupled to the scalar chain. 
}
\end{figure}

\begin{figure}[htb]
\epsfxsize=3.2in
\epsffile{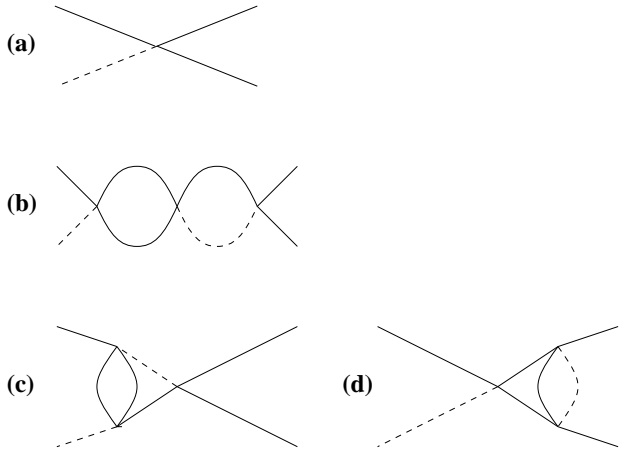}

\vspace*{0.5cm}

\caption{The interaction vertex function in the first and 
third order perturbation theory. (a) is the original interaction; (b), (c), and (d)
are the third order corrections. (b) and (c) are parquet diagrams, producing the
leading logarithmic terms.}
\end{figure}

\begin{figure}[htb]
\epsfxsize=3.2in
\epsffile{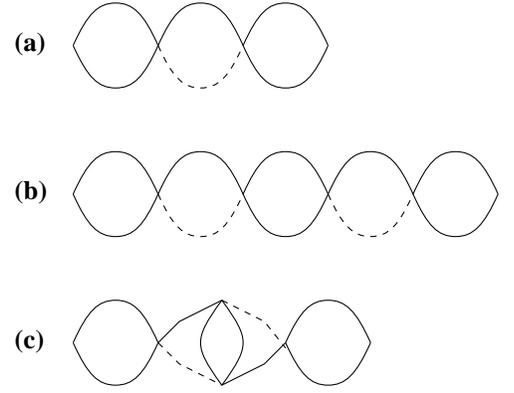}

\vspace*{0.5cm}

\caption{The total spin susceptibility in the second and fourth order perturbation theory.
(a) describes the second order correction,  giving rise to a logarithmic term; (b) and (c) 
are the leading order singular contributions in the fourth order 
perturbation theory, producing
the squared logarithmic corrections.  }
\end{figure}

\begin{figure}[htb]
\epsfxsize=3.2in
\epsffile{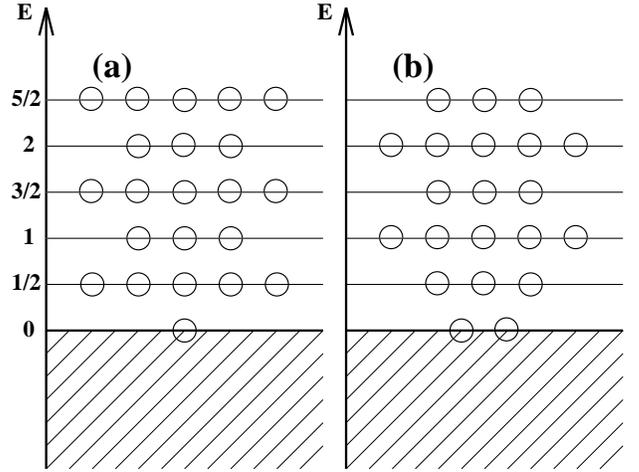}

\vspace*{0.5cm}

\caption{Single particle spectra for the two sectors present at the
non-Fermi liquid fixed point of the two channel Kondo model. (a) corresponds
to sector I and (b) to sector II.
 The energies are given in units of
$\frac{\pi v_{\rm F}}{l}$.}
\end{figure}

\begin{figure}[htb]
\epsfxsize=3.2in
\epsffile{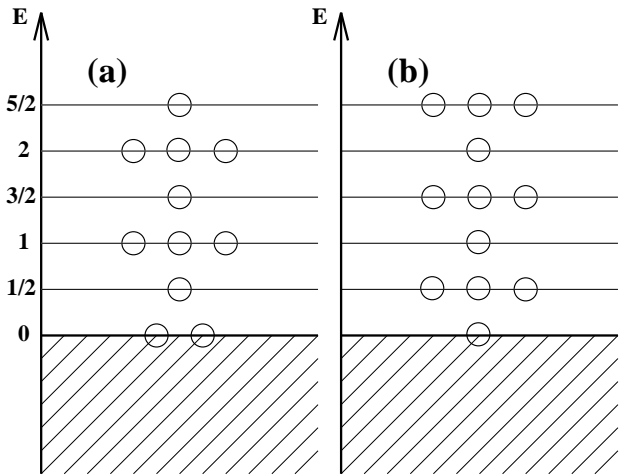}

\vspace*{0.5cm}

\caption{Single particle spectra of the compactified models.
(a) for PBC in the vector part and APBC in the scalar part 
(corresponding to $N$ odd)
and (b)
for APBC in the vector part and PBC in the scalar part 
(corresponding to $N$ even).
 The energies are given in units of
$\frac{\pi v_{\rm F}}{l}$.}
\end{figure}

\end{document}